\documentclass[runningheads]{llncs}

\usepackage[T1]{fontenc}
\usepackage{graphicx}
\usepackage{xcolor}
\usepackage{listings}
\usepackage{subcaption}
\usepackage{booktabs}
\usepackage{amssymb}
\usepackage{mathtools}
\usepackage{ebproof}
\usepackage{mathpartir}
\usepackage{microtype}

\providecommand{\Description}[1]{}

\newif\ifdraft%
\drafttrue%

\definecolor{codebg}{gray}{0.92}

\lstdefinelanguage{hazel}{
  columns=fullflexible,
  keepspaces=true,
  basicstyle=\ttfamily\color{black},
  keywords={
    let, in, case, end, type,
    eval, stop, hide, step, debug,
    fun,
    if, then, else,
  },
  keywordstyle=\bfseries\color{black},
  commentstyle=\color{gray},
}

\newcommand{\code}[1]{{\fboxsep=1.5pt\colorbox{codebg}{\texttt{\small #1}}}}

\DeclareMathSymbol{\Gamma}{\mathalpha}{operators}{0}
\DeclareMathSymbol{\Delta}{\mathalpha}{operators}{1}
\DeclareMathSymbol{\Theta}{\mathalpha}{operators}{2}
\DeclareMathSymbol{\Lambda}{\mathalpha}{operators}{3}
\DeclareMathSymbol{\Xi}{\mathalpha}{operators}{4}
\DeclareMathSymbol{\Pi}{\mathalpha}{operators}{5}
\DeclareMathSymbol{\Sigma}{\mathalpha}{operators}{6}
\DeclareMathSymbol{\Upsilon}{\mathalpha}{operators}{7}
\DeclareMathSymbol{\Phi}{\mathalpha}{operators}{8}
\DeclareMathSymbol{\Psi}{\mathalpha}{operators}{9}
\DeclareMathSymbol{\Omega}{\mathalpha}{operators}{10}

\DeclareMathOperator{\DefPat}{Pat}
\DeclareMathOperator{\DefExp}{Exp}

\DeclareMathOperator{\DefCtx}{Ctx}
\DeclareMathOperator{\DefAct}{Act}
\DeclareMathOperator{\DefGas}{Gas}
\DeclareMathOperator{\DefLvl}{Priority}
\DeclareMathOperator{\DefFilter}{Filter}
\DeclareMathOperator{\DefTyp}{Typ}

\DeclareMathOperator{\KeywordFilter}{debug}

\newcommand{\PatExpr}{\mathtt{\$e}}
\newcommand{\PatValue}{\mathtt{\$v}}

\newcommand{\ActSkip}{\mathsf{skip}}
\newcommand{\ActStep}{\mathsf{step}}

\newcommand{\GasOne}{\mathsf{one}}
\newcommand{\GasAll}{\mathsf{all}}

\newcommand{\Nat}[1]{#1}
\newcommand{\Lam}[2]{\lambda #1 . #2}
\newcommand{\Fix}[2]{\mu #1 . #2}

\newcommand{\Filter}[2]{\KeywordFilter{} #1~\mathsf{in}~#2}

\newcommand{\Residue}[4]{\prescript{#1}{#2}{\langle} #4 \rangle^{#3}}

\newcommand{\FSubst}[3]{[#1/#2] #3}
\DeclareMathOperator{\Strip}{strip}
\DeclareMathOperator{\Decay}{decay}

\newcommand{\jbox}[1]{\par\medskip\noindent\fbox{#1}\;}

\newcommand{\Mark}{\circ}

\newcommand{\Value}[1]{#1~\mathsf{value}}

\newcommand{\Decompose}[3]{#1 \Rightarrow #2 \{ #3 \}}
\newcommand{\Compose}[3]{#1 \Leftarrow #2 \{ #3 \}}

\newcommand{\Matches}[2]{#1 \mathrel{\mathop{\triangleright}} #2}
\newcommand{\DoesNotMatch}[2]{#1 \mathrel{\mathop{\not\triangleright}} #2}

\newcommand{\InstructPAGL}[6]{(#1, #2, #3, #4) \vdash #5 \rightsquigarrow #6}

\newcommand{\Analyze}[3]{#1 \vdash #2 \dashv #3}

\newcommand{\Transition}[2]{#1 \rightarrow #2}

\newcommand{\Step}[3]{#1 \twoheadrightarrow^{#3} #2}
\newcommand{\JustStep}[3]{#1 \mapsto^{#3} #2}
\newcommand{\JustStepStar}[2]{#1 \mapsto^{*} #2}

\newcommand{\Strippable}[1]{#1~\mathsf{strippable}}

\newcommand{\IsResidue}[1]{#1~\mathsf{is residue}}
\newcommand{\Optimize}[2]{#1 \rightarrow_{opt} #2}

\newcommand{\Arrow}[2]{#1 \Rightarrow #2}
\newcommand{\Natural}{\mathbb{N}}
\newcommand{\TypeEntails}[3]{#1 \vdash #2 : #3}
\newcommand{\TypeContains}[3]{#1 \ni #2 : #3}
\newcommand{\TypeExtends}[3]{#1 , #2 : #3}

\usepackage{hyperref}

\begin{document}

\title{Practical Algebraic Stepping with Scoped Filters}


\author{Haoxiang Fei \and Matthew Keenan \and Cyrus Omar}
\authorrunning{H. Fei et al.}
\institute{University of Michigan, Ann Arbor, MI, USA\\
  \email{\{fettes,mckeenan,comar\}@umich.edu}}




\maketitle

\begin{abstract}
  Algebraic steppers help students learn functional programming by
  displaying evaluation as a sequence of small-step reductions, but
  even simple programs produce long traces in which key ideas are
  buried under mundane reductions. This paper presents the
  \emph{filtered stepper calculus}, a formal framework that gives
  users scoped, pattern-based control over which reduction steps are
  shown or hidden. Users annotate programs with lightweight filter
  expressions that match on the structure of redexes. Filters compose
  via lexical scoping so that inner filters override outer ones. We
  prove preservation, progress, and a simulation theorem establishing
  that the filtered stepper agrees with the underlying unfiltered
  semantics, and mechanize all proofs in Agda. We implement the
  calculus in the Hazel live programming environment, including its support for stepping programs with holes and type errors. To do so, we reconcile
  Hazel's internal environment-based evaluator with the
  substitution-based presentation expected in the classroom. We 
  deploy the system in a university programming languages course. Our
  evaluation shows that students adopt the stepper organically, though more advanced uses of filters may require further instruction, and
  that instructors can use filters to craft focused traces for
  use in lectures.
    
\end{abstract}

\section{Introduction}\label{sec:intro}

Substitution-based equational reasoning is a cornerstone of functional
programming education: students learn to evaluate programs by
repeatedly rewriting redexes, producing a trace of steps that serves
as a proof of the final result. Hazel~\cite{omar2017hazelnut} is a
live functional programming environment that has been adopted for use
in an upper-level programming languages course in which this style of
reasoning is taught. A natural pedagogical tool in this setting is an
\emph{algebraic stepper}---a ``debugger'' that presents each small-step
reduction as a justified equational step, mirroring what students see
in typeset or handwritten lectures and assignments.

\begin{figure}[h]
  \begin{center}
    \begin{subfigure}[t]{0.52\textwidth}
      \centering
      \frame{\includegraphics[width=\textwidth]{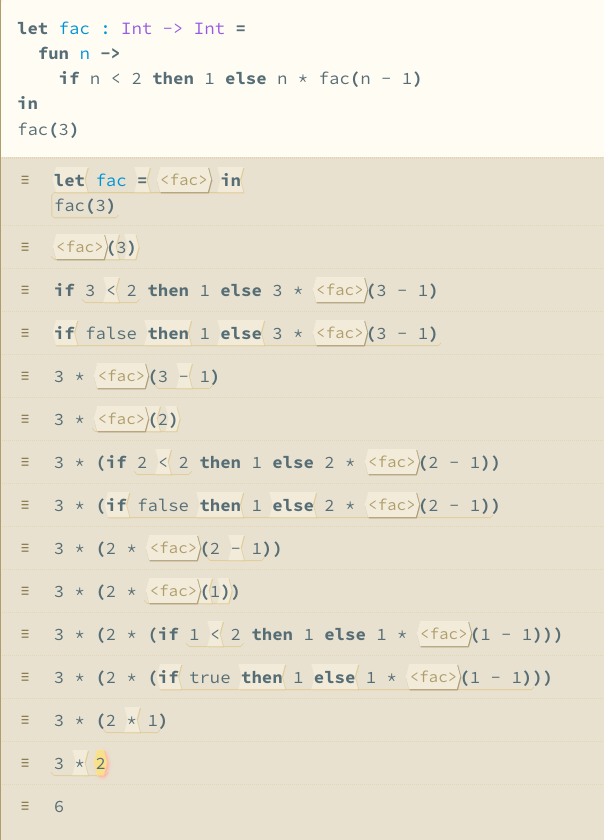}}
      \Description{Unfiltered stepper trace for a factorial program, showing many small-step equations from the function definition to the final result 6.}
      \caption{Program trace without filters.}
      \label{fig:intro-fac-unfiltered}
    \end{subfigure}
    \quad
    \begin{subfigure}[t]{0.43\textwidth}
      \centering
      \frame{\includegraphics[width=\textwidth]{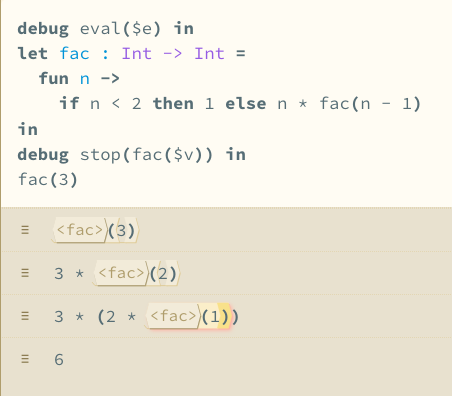}}
      \Description{Filtered stepper trace for the same factorial program, showing only a few key steps and the final result 6.}
      \caption{Program trace with filters that stop at applications of the function to a value.}
      \label{fig:intro-fac-filtered}
    \end{subfigure}
  \end{center}
  \caption{Stepper traces for a factorial program. Filters reduce a long
    sequence of small steps to just the relevant reductions.}
  \label{fig:intro-fac}
\end{figure}

However, even moderately sized programs produce long traces. Computing
\code{fac(3)}, for instance, already requires over a dozen small steps
when every arithmetic operation and substitution is shown
(Figure~\ref{fig:intro-fac-unfiltered}), and realistic classroom
examples (recursive list traversals, higher-order functions, pattern
matching) generate traces that are far longer. Most of these steps
are mundane: they perform arithmetic that the student already
understands, or substitute values into bindings that are not relevant
to the concept being taught. The sheer volume obscures the key ideas
the instructor wants to highlight and makes the stepper tedious to use
in lectures and homework alike.

Existing tools offer limited help with this problem. The Racket algebraic stepper~\cite{clements2001stepper}
faithfully presents small-step reductions but provides no mechanism
for the user to select which \emph{kinds} of steps appear in the
trace. Conditional breakpoints, as found in most debuggers, allow
stopping at particular locations or when a variable satisfies a
predicate, but they cannot express structural conditions such as
``pause whenever a recursive call to \texttt{map} is about to be
applied to two values.''

We address this gap with a \emph{scoped filter system} that gives
users pattern-based control over which reduction steps are shown or
hidden. Users write lightweight filter annotations, embedded directly
in the program text, that specify patterns over expressions. A filter
can \emph{hide} all steps matching a pattern, \emph{show} (stop at) steps
matching a pattern, or apply these effects to the entire evaluation of
a matched sub-expression rather than a single step. Filters compose
naturally. An inner filter can override an outer one through lexical
nesting, and a filter attached to a function body travels with the
closure, taking effect wherever the function is called.

Returning to the factorial example, suppose the instructor wants to
show only the recursive calls and hide the routine arithmetic. Adding
\code{debug eval(\$e)} hides every step, while \code{debug stop(fac(\$v))}
overrides this on applications of \code{fac} to a value. The resulting
trace (Figure~\ref{fig:intro-fac-filtered}) shows only the key
recursive calls and the final result.

We make the following contributions:
\begin{itemize}
  \item We present the \emph{Filtered Stepper Calculus}, a formal
    semantics for scoped, pattern-based stepping control. We introduce
    its filter primitives by example in Section~\ref{sec:examples} and
    formalize them in Section~\ref{sec:filter}, with preservation,
    progress, and simulation all mechanized in Agda.
  \item We implement the filtered stepper in Hazel
    (Section~\ref{sec:implementation}), reconciling Hazel's internal
    environment-based big-step evaluator with the substitution-based
    presentation expected by students.
  \item We deploy the stepper in a university programming languages
    course (Section~\ref{sec:evaluation}) and report on the
    instructor's experience in lecture, student usage data from
    assignments, and end-of-term survey responses.
\end{itemize}

Section~\ref{sec:related} situates the work in related literature, and
Section~\ref{sec:discussion} concludes.

\paragraph{Supplemental material.}
Accompanying this submission are the Agda mechanization, the OCaml
source and pre-built binary of the reference implementation, and a
pre-built copy of the Hazel stepper.


\section{Hazel Stepper by Example}\label{sec:examples}

Section~\ref{sec:intro} showed how a single filter can collapse a
factorial trace to its key steps. We now use a richer example,
mapping a function over a list, to introduce the full set of filter
primitives and show how they compose.

\begin{figure}
  \centering
  \begin{subfigure}[T]{0.40\linewidth}
    \frame{\includegraphics[scale=0.34]{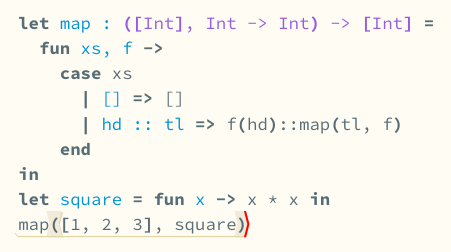}}
    \Description{Hazel code defining a list map function, a square function, and a call to map over [1, 2, 3].}
    \caption{The map example. Evaluates to \code{[1, 4, 9]}.}
    \label{fig:hazel-map}
  \end{subfigure}
  \hfill
  \begin{subfigure}[T]{0.50\linewidth}
    \frame{\includegraphics[scale=0.37]{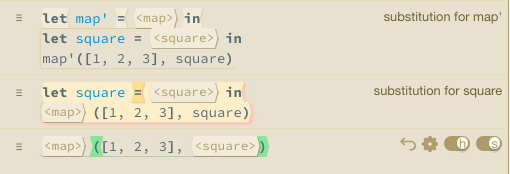}}
    \Description{Partial stepper trace for the map example, stopped at map([1, 2, 3], square), with highlighted next-step location and substitution annotations.}
    \caption{Partial stepper trace for the map example, stopped at \code{map([1, 2, 3], square)}. Each line is one small-step transition. The green highlight at the bottom marks where the next step will be taken.}
    \label{fig:full-trace}
  \end{subfigure}
  \caption{The stepper used on a list map example.}
\end{figure}

Consider the Hazel program in \autoref{fig:hazel-map}, which applies a
\code{square} function to each element of the list \code{[1, 2, 3]}
using \code{map}. To understand how this program evaluates to
\code{[1, 4, 9]}, a user can invoke the Hazel stepper. The stepper
highlights every reducible expression (redex) in the program. The user
clicks on a highlighted redex to reduce it, and the expression is
rewritten in place. A history panel lets the user expand the full
sequence of steps taken so far. \autoref{fig:full-trace} shows the
partial trace for this program: each line corresponds to one
reduction step, and the green highlight at the bottom marks the
available redexes for the next step.

The factorial traces in \autoref{fig:intro-fac} illustrate why showing
every step can be distracting: routine substitutions and arithmetic
quickly obscure the reductions that matter. The partial trace in
\autoref{fig:full-trace} shows the same issue arising for \code{map}.
To address this, the Hazel stepper provides \emph{filter} annotations
that the user embeds in the program text to control which steps appear
in the trace.

Each filter takes a \emph{pattern} that selects expressions by
structure. Patterns use the same syntax as Hazel expressions, so
every Hazel expression is also a valid pattern. Two additional forms
are provided: \code{\$e}, which matches any expression, and
\code{\$v}, which matches any value. For example,
\code{map(\$v, \$v)} matches any application of \code{map} to two
values. We give the formal pattern-matching rules in
Section~\ref{sec:filter}. We provide four filter constructs:
\begin{itemize}
  \item \code{debug hide(\textit{pat})}: hide \emph{one} matching step.
  \item \code{debug stop(\textit{pat})}: show \emph{one} matching step.
  \item \code{debug eval(\textit{pat})}: hide \emph{all} evaluation steps for a matching sub-expression.
  \item \code{debug step(\textit{pat})}: show \emph{all} evaluation steps for a matching sub-expression.
\end{itemize}
The first two (\code{hide}/\code{stop}) affect a single step, while the
latter two (\code{eval}/\code{step}) affect the entire evaluation of a
matching sub-expression.

\begin{figure}
  \centering
  \begin{subfigure}[T]{0.40\linewidth}
    \frame{\includegraphics[scale=0.4]{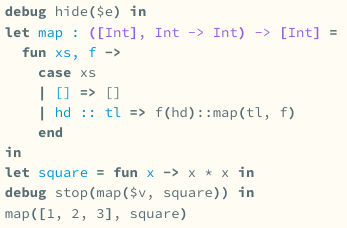}}
    \Description{Map example annotated with debug eval and debug stop filters to focus on recursive calls.}
    \caption{The code from \autoref{fig:hazel-map} with added \code{debug hide} and \code{debug stop} filters.}
    \label{fig:hide-stop-map}
  \end{subfigure}
  \hfill
  \begin{subfigure}[T]{0.50\linewidth}
    \frame{\includegraphics[scale=0.4]{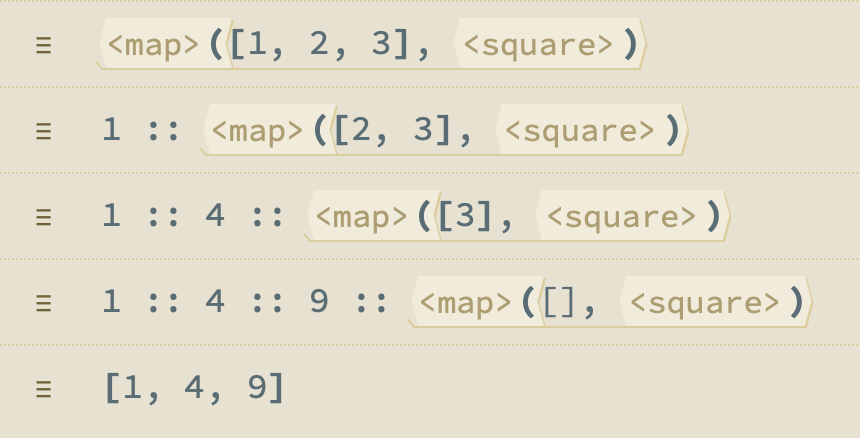}}
    \Description{Filtered stepper output showing only recursive map calls and the resulting list [1, 4, 9].}
    \caption{Filtered trace. Each line shows one recursive \code{map} call: the list shrinks by one element while evaluated results accumulate on the left.}
    \label{fig:recursive-trace}
  \end{subfigure}
  \caption{A filtered stepper for showing recursive calls in the list map example.}
\end{figure}

To illustrate, suppose we only want to see the recursive calls to
\code{map}. In \autoref{fig:hide-stop-map} we add two filters:
\code{debug hide(\$e)} hides every step, since \code{\$e} re-matches
at every instrumentation pass, and \code{debug stop(map(\$v, \$v))}
overrides the preceding hide to show any step taken from an
expression containing \code{map} applied to two values. Since
\code{debug stop} is the inner of the two filters, it takes
priority. The resulting trace (\autoref{fig:recursive-trace}) shows
only the recursive \code{map} calls.

\begin{figure}
  \centering
  \begin{subfigure}[T]{0.40\linewidth}
    \frame{\includegraphics[scale=0.30]{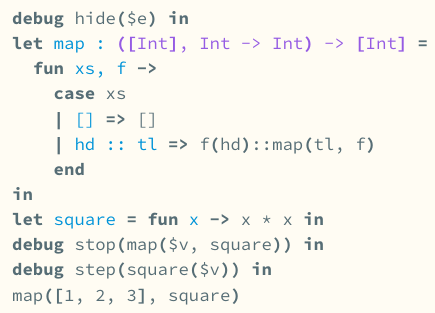}}
    \Description{Map example with eval and step filters, including debug mode for square.}
    \caption{The code from \autoref{fig:hazel-map} with added \code{debug hide}, \code{debug stop}, and \code{debug step} filters.}
    \label{fig:debug-map}
  \end{subfigure}
  \hfill
  \begin{subfigure}[T]{0.56\linewidth}
    \frame{\includegraphics[scale=0.30]{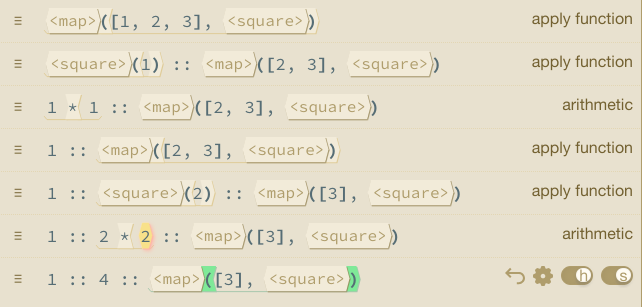}}
    \Description{Filtered trace showing recursive map steps and explicit square evaluations on list elements.}
    \caption{Filtered trace with \code{debug step}: recursive \code{map} calls are visible (as before), and each \code{square} application is expanded to show its intermediate steps.}
    \label{fig:debug-trace}
  \end{subfigure}
  \caption{A filtered stepper that demonstrates debug mode.}
\end{figure}

Alternatively, we might want to see not just which recursive calls
\code{map} makes, but also how each application of \code{square}
evaluates. In \autoref{fig:debug-map} we add a \code{debug step}
filter to achieve this. The result (\autoref{fig:debug-trace}) now
includes not only the recursive \code{map} calls, but also the
intermediate multiplications inside each \code{square} call. For
instance, the third and sixth lines show \code{<square>(1)} and
\code{<square>(2)} being expanded into their constituent steps.

Note that \code{debug step} here overrides the earlier \code{debug hide}
and \code{debug stop}. Intuitively, the innermost filter surrounding
the next reduction step takes priority, a form of lexical scoping.
In the next section, we formalize these filter primitives and their
interactions as the \emph{Filtered Stepper Calculus}.

\section{Filtered Stepper Calculus}\label{sec:filter}

We now formalize the filter system as the \emph{filtered stepper
calculus}. The key idea is that filter annotations in the source
program are compiled into \emph{residues}, internal markers that
record what action to take when a step occurs at a given
sub-expression. At each step, the stepper selects the
highest-priority residue surrounding the current redex to decide
whether to show the step or skip it silently. We present the syntax,
contextual dynamics, static semantics, and metatheoretic properties.

\subsection{Syntax}

The four filter constructs introduced in
Section~\ref{sec:examples} each desugar into a \(\Filter{f}{e}\)
expression parameterized by a triple \(f = (p, a, g)\) of a pattern,
an action, and a gas, as shown in
Table~\ref{tbl:filter-syntax-sugar}.

\begin{table}[htpb]
  \centering
  \caption{Desugaring of filter constructs. Each filter is a triple of
    a pattern~\(p\), an action~\(a\), and a gas~\(g\).}\label{tbl:filter-syntax-sugar}
  \setlength{\tabcolsep}{8pt}
  \renewcommand{\arraystretch}{1.3}
  \begin{tabular}{|l|l|}
    \hline
    \textbf{Hazel code} & \textbf{Desugared form} \\
    \hline
    \verb|debug hide(p) in e| & \(\Filter{(p, \ActSkip, \GasOne)}{e}\) \\
    \verb|debug eval(p) in e| & \(\Filter{(p, \ActSkip, \GasAll)}{e}\) \\
    \verb|debug stop(p) in e| & \(\Filter{(p, \ActStep, \GasOne)}{e}\) \\
    \verb|debug step(p) in e| & \(\Filter{(p, \ActStep, \GasAll)}{e}\) \\
    \hline
  \end{tabular}
\end{table}

The \emph{action} \(a\) determines what happens when a step matches:
\(\ActStep\) shows the step to the user, \(\ActSkip\) hides it. The
\emph{gas} \(g\) determines the lifetime of the effect: \(\GasOne\)
applies to a single step, \(\GasAll\) applies to all steps in the
evaluation of the matched sub-expression.

The full syntax (Fig.~\ref{fig:filter-syntax}) extends the untyped
lambda calculus with natural numbers, addition, and a fixpoint operator
\(\Fix{x}{e}\) for recursive definitions. Two expression forms are
added: \(\Filter{f}{e}\), which the user writes, and
\(\Residue{a}{g}{l}{e}\), which is generated internally by the
stepper. A residue marks a sub-expression where a filter has already
matched, recording the action~\(a\), gas~\(g\), and a
priority~\(l\), a natural number reflecting how recently the
filter was encountered, so that inner filters override outer ones.

Patterns mirror the expression grammar but add two wildcard forms:
\(\PatExpr\), which matches any expression, and \(\PatValue\), which
matches any value.

\begin{figure}[h]
  \begin{equation*}
    \begin{array}{rcl}
      \DefAct a    &\Coloneqq& \ActSkip \mid \ActStep \\
      \DefGas g    &\Coloneqq& \GasOne \mid \GasAll \\
      \DefLvl l    &\Coloneqq& \mathbb{N} \\
      \DefFilter f &\Coloneqq& (p, a, g) \\
      \DefExp e    &\Coloneqq& x \mid e(e) \mid \Lam{x}{e} \mid \Fix{x}{e} \mid e + e \mid \Nat{n} \\
                   &\mid     & \Filter{f}{e} \mid \Residue{a}{g}{l}{e} \\
      \DefPat p    &\Coloneqq& x \mid p(p) \mid \Lam{x}{p} \mid p + p \mid \Nat{n} \mid \PatExpr \mid \PatValue \\
      \DefCtx \mathcal{E} &\Coloneqq& \Mark \mid \mathcal{E}(e) \mid e(\mathcal{E}) \mid \mathcal{E} + e \mid e + \mathcal{E} \\
                   &\mid     & \Filter{f}{\mathcal{E}} \mid \Residue{a}{g}{l}{\mathcal{E}} \\
    \end{array}
  \end{equation*}
  \caption{Syntax of the Filtered Stepper Calculus. Here, $n$ ranges over numbers, $x$ ranges over variables, and $l$ ranges over priorities, which are natural numbers. Evaluation contexts $\mathcal{E}$ have exactly one hole, denoted $\Mark$.}
  \label{fig:filter-syntax}
\end{figure}

Values are lambda abstractions and natural-number literals. Fixpoints
are not values: they unroll during evaluation. Filter and residue
wrappers around a value are not values either. They are eliminated by
the instruction transition relation introduced later in this section.

\jbox{\(\Value e\)} Expression \(e\) is a value.
\begin{mathpar}
  \inferrule[V-Lam]{ }{
    \Value{\Lam{x}{e}}
  } \qquad
  \inferrule[V-Nat]{ }{
    \Value{\Nat{n}}
  }
\end{mathpar}

\subsection{Contextual Dynamics by Example}

Each visible step in the stepper is the result of a multi-phase
pipeline: \emph{instrumentation}, \emph{optimization},
\emph{decomposition}, \emph{action selection}, \emph{decay,
transition, and composition}, and \emph{user interaction}. We
introduce each phase by tracing one example step of a small program
through the pipeline, then present the formal rules.

Consider the program

\begin{lstlisting}[language=hazel,xleftmargin=1in]
debug eval(1 + 2 + 3 + 4) in
debug stop(3 + 3 + 4) in
1 + 2 + 3 + 4
\end{lstlisting}
\noindent which sums the numbers from 1 to 4 left-associatively. Before
stepping, the stepper implicitly wraps the program in a default
\code{debug stop(\$e)} so that every step is shown unless an
explicit filter says otherwise. Together with the two explicit
filters, there are then three active filters. In lexical order from
outermost to innermost, they receive priorities 0, 1, and 2.
Outermost gets the lowest priority so that inner filters can
override outer ones during action selection, realizing the
lexical-scoping intuition that the filter ``closest to'' a redex has
the final say. Rule names below refer to the formal rules introduced
later in this section.

\paragraph{Instrumentation.}
The stepper traverses the expression collecting active filters, and
wraps each sub-expression that matches a filter's pattern in a
\emph{residue} recording the filter's action, gas, and priority. In
our example, every sub-expression matches the default
\code{debug stop(\$e)} (\textsc{I-Add-Y}); the entire sum
additionally matches \code{debug eval(1 + 2 + 3 + 4)}. Values are
not wrapped (\textsc{I-V}), since values do not step. The third filter, \code{debug stop(3 + 3 + 4)}, does not yet match anything: the sub-expression \code{3 + 3 + 4} only appears after one reduction step, so it contributes no residue (\textsc{I-Add-N}). The result, with the surrounding filter prefix elided for brevity, is:
\begin{lstlisting}[language=hazel]
@\(
\prescript{\ActStep}{\GasOne}{\langle}
\prescript{\ActSkip}{\GasAll}{\langle}
\prescript{\ActStep}{\GasOne}{\langle}
\prescript{\ActStep}{\GasOne}{\langle}
\)@1 + 2@\({\rangle}^{0}\)@ + 3@\({\rangle}^{0}\)@ + 4@\(\rangle^{1}
\rangle^{0}\)@
\end{lstlisting}

\paragraph{Optimization.}
Adjacent residues are collapsed, keeping only the highest-priority
one. The pass is observationally invisible, preserving the action
that selection would compute at every redex, but is essential for
keeping the stepper tractable: without it, residue chains can grow
exponentially in the depth of nested filters within a single
instrumentation pass, and persist across steps so that the chain
length accumulates over the history of stepping. We discuss the
asymptotic effect in Section~\ref{sec:impl-perf}. In our example,
the two outermost residues are adjacent;
\textsc{O-Residue-Inner} drops the outer
\(\Residue{\ActStep}{\GasOne}{0}{\cdot}\) in favor of the
higher-priority inner \(\Residue{\ActSkip}{\GasAll}{1}{\cdot}\):
\begin{lstlisting}[language=hazel]
@\(
\prescript{\ActSkip}{\GasAll}{\langle}
\prescript{\ActStep}{\GasOne}{\langle}
\prescript{\ActStep}{\GasOne}{\langle}
\)@1 + 2@\({\rangle}^{0}\)@ + 3@\({\rangle}^{0}\)@ + 4@\(
\rangle^{1}
\)@
\end{lstlisting}

\paragraph{Decomposition.}
The expression is split into an evaluation context \(\mathcal{E}\)
and a selected redex \(e_0\); the relation is non-deterministic,
as the stepper offers the user a choice of redexes.
However, this choice is only meaningful when the step will be shown to the user: if the action selection that follows yields \(\ActSkip\), then the stepper picks one redex deterministically (in practice, the first) and continues.
Here we select \code{1 + 2} as the redex, yielding the
evaluation context:
\begin{lstlisting}[language=hazel]
@\(
\prescript{\ActSkip}{\GasAll}{\langle}
\prescript{\ActStep}{\GasOne}{\langle}
\prescript{\ActStep}{\GasOne}{\langle}
\circ{\rangle}^{0}\)@ + 3@\({\rangle}^{0}\)@ + 4@\(
\rangle^{1}
\)@
\end{lstlisting}

\paragraph{Action selection.}
Walking outward through the residues surrounding the redex in
\(\mathcal{E}\), the highest-priority action wins. In our example,
the two inner residues carry \(\ActStep\) at priority 0 (no
override of the starting \(\ActStep, 0\)), while the outer residue
carries \(\ActSkip\) at priority 1 and takes over. The selected
action is \(\ActSkip\).

\begin{figure}[h]
  \centering
  \frame{\includegraphics[width=0.55\textwidth]{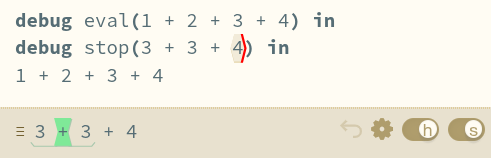}}
  \Description{Stepper UI showing the expression 1 + 2 + 3 + 4 with a highlighted subexpression selection and control buttons.}
  \caption{Stepper UI at the user-selection step.}\label{fig:simple_example_final_ui}
\end{figure}

\paragraph{Decay, transition, and composition.}
The redex is reduced by an instruction transition, the surrounding
context is \emph{decayed} to remove one-shot residues (\(\GasOne\))
while preserving persistent ones (\(\GasAll\)), and the transition
result is plugged back into the decayed context. In our example,
the redex \code{1 + 2} transitions to \code{3} (\textsc{T-Add}),
the persistent \(\GasAll\) residue at priority 1 survives, and
composition yields:
\begin{lstlisting}[language=hazel]
@\(\prescript{\ActSkip}{\GasAll}{\langle}\)@3 + 3 + 4@\(\rangle^{1}\)@
\end{lstlisting}

\paragraph{User interaction and repetition.}
If the selected action is \(\ActStep\) and the redex is not a filter or residue elimination, the stepper pauses and presents the step to the user. Otherwise it skips silently and continues from instrumentation. Because the action in this round was \(\ActSkip\), the stepper does not pause---it recurses. Re-instrumenting \(\Residue{\ActSkip}{\GasAll}{1}{3 + 3 + 4}\), the third filter \code{debug stop(3 + 3 + 4)} now matches the inner sum and inserts a residue at priority 2. Decomposing to select \code{3 + 3} as the redex yields:
\begin{lstlisting}[language=hazel]
@\(
\prescript{\ActSkip}{\GasAll}{\langle}
\prescript{\ActStep}{\GasOne}{\langle}
\prescript{\ActStep}{\GasOne}{\langle}
\prescript{\ActStep}{\GasOne}{\langle}
\circ
\rangle^{0}
\)@ + 4@\(
\rangle^{2}
\rangle^{0}
\rangle^{1}
\)@
\end{lstlisting}
The priority-2 residue carries action \(\ActStep\) and overrides
the outer \(\ActSkip\) at priority 1. Action selection therefore
yields \(\ActStep\), and since the redex is not strippable the
stepper pauses for user input
(Figure~\ref{fig:simple_example_final_ui}). In general, several
silent skips, possibly in different parts of the expression, may
intervene between two user-visible steps.

\subsection{Semantics for Contextual Dynamics}\label{sec:dynamics}

We now present the formal rules that define each phase, building
up to the top-level stepping judgment.

\subsubsection{Instrumentation.} The instrumentation judgment traverses
an expression, testing each sub-expression against the current filter's
pattern using the matching judgment \(\Matches{p}{e}\). When a match is found (the ``-Y'' rules), a residue is
inserted. Otherwise (the ``-N'' rules), the sub-expression is left
unchanged. When a nested filter is encountered, its priority is
incremented so that inner filters take precedence.

\jbox{\(\InstructPAGL{p}{a}{g}{l}{e}{e'}\)} \(e\) is dynamically instrumented to \(e'\).
\begin{mathpar}
  \inferrule[I-V]{
    \Value{v}
  }{
    \InstructPAGL{p}{a}{g}{l}{v}{v}
  } \qquad
  \inferrule[I-Var]{
  }{
    \InstructPAGL{p}{a}{g}{l}{x}{x}
  } \\
  \inferrule[I-Filter]{
    \InstructPAGL{p_0}{a_0}{g_0}{l_0}{e_0}{e} \\
    \InstructPAGL{p}{a}{g}{l_0 + 1}{e}{e'}
  }{
    \InstructPAGL{p_0}{a_0}{g_0}{l_0}{\Filter{(p, a, g)}{e_0}}{\Filter{(p, a, g)}{e'}}
  } \\
  \inferrule[I-Residue]{
    \InstructPAGL{p_0}{a_0}{g_0}{l_0}{e_0}{e} \\
  }{
    \InstructPAGL{p_0}{a_0}{g_0}{l_0}{\Residue{a}{g}{l}{e_0}}{\Residue{a}{g}{l}{e}}
  } \\
  \inferrule[I-Fix]{
    \
  }{
    \InstructPAGL{p}{a}{g}{l}{\Fix{x}{e}}{\Fix{x}{e}}
  } \\
  \inferrule[I-Ap-Y]{
    \InstructPAGL{p}{a}{g}{l}{e_1}{e_1'} \\
    \InstructPAGL{p}{a}{g}{l}{e_2}{e_2'} \\
    \Matches{p}{e_1(e_2)}
  }{
    \InstructPAGL{p}{a}{g}{l}{e_1(e_2)}{\Residue{a}{g}{l}{e_1'(e_2')}}
  } \\
  \inferrule[I-Ap-N]{
    \InstructPAGL{p}{a}{g}{l}{e_1}{e_1'} \\
    \InstructPAGL{p}{a}{g}{l}{e_2}{e_2'} \\
    \DoesNotMatch{p}{e_1(e_2)}
  }{
    \InstructPAGL{p}{a}{g}{l}{e_1(e_2)}{e_1'(e_2')}
  } \\
  \inferrule[I-Add-Y]{
    \InstructPAGL{p}{a}{g}{l}{e_1}{e_1'} \\
    \InstructPAGL{p}{a}{g}{l}{e_2}{e_2'} \\
    \Matches{p}{e_1 + e_2}
  }{
    \InstructPAGL{p}{a}{g}{l}{e_1 + e_2}{\Residue{a}{g}{l}{e_1' + e_2'}}
  } \\
  \inferrule[I-Add-N]{
    \InstructPAGL{p}{a}{g}{l}{e_1}{e_1'} \\
    \InstructPAGL{p}{a}{g}{l}{e_2}{e_2'} \\
    \DoesNotMatch{p}{e_1 + e_2}
  }{
    \InstructPAGL{p}{a}{g}{l}{e_1 + e_2}{e_1' + e_2'}
  } \\
\end{mathpar}

Two rules deserve a brief comment. \textsc{I-Filter} traverses the
body of a nested filter twice: once with the surrounding filter
\((p_0, a_0, g_0, l_0)\), and once more with the newly-entered filter
at priority \(l_0 + 1\). Equivalently, every filter currently in
scope gets its own instrumentation pass over each sub-expression it
covers, and the priority assigned to a residue records which pass
inserted it. This is what makes inner filters override outer ones
at the same redex during action selection. A side effect is that
the residue count at a single sub-expression can be exponential in
the nesting depth of filters before optimization, since each level
of nesting potentially re-walks the residues already added. The
optimization pass below collapses these chains, restoring an $O(k)$
residue count per node for $k$ filters in scope.

\textsc{I-Fix} does not recurse into the body of a fixpoint. The
reason is that \textsc{T-Fix} unrolls a fixpoint by substituting it
into its own body. Substitution preserves residues
(Appendix~\ref{sec:standard_rules}), so any residues placed inside
the body before unrolling would be replicated into every copy
produced by the unrolling. Re-instrumenting after the unrolling
step places residues at the correct sub-expressions in the
expanded term instead.

Instrumentation is the cost center of a step: each step re-instruments
the entire subtree, and because residues persist across steps, these
re-instruments leave more and more residues in the expression. The
optimization pass below exists to keep that work from compounding across
steps.

\subsubsection{Pattern matching.} The matching judgment determines
when a sub-expression's structure conforms to a filter's pattern.
Because filter and residue wrappers are purely instrumental, matching
ignores them by comparing expression bodies modulo the stripping
operation defined below. In the rules \textsc{M-Lam} and
\textsc{M-Fix}, the equivalence \(\Strip{e_1} \equiv \Strip{e_2}\)
is up to \(\alpha\)-equivalence. The Agda mechanization enforces
this implicitly by using de Bruijn indices throughout.

\jbox{\(\Matches{p}{e}\)} Pattern \(p\) matches expression \(e\).
\begin{mathpar}
  \inferrule[M-All]{\ }{
    \Matches{\$e}{e}
  } \qquad
  \inferrule[M-Val]{\Value{v}}{
    \Matches{\$v}{v}
  } \qquad
  \inferrule[M-Nat]{
    \
  }{
    \Matches{\Nat{n}}{\Nat{n}}
  } \\
  \inferrule[M-Lam]{
    \Strip{e_1} \equiv \Strip{e_2}
  }{
    \Matches{\Lam{x_1}{e_1}}{\Lam{x_2}{e_2}}
  } \qquad
  \inferrule[M-Fix]{
    \Strip{e_1} \equiv \Strip{e_2}
  }{
    \Matches{\Fix{x_1}{e_1}}{\Fix{x_2}{e_2}}
  } \\
  \inferrule[M-Ap]{
    \Matches{p_1}{e_1} \\
    \Matches{p_2}{e_2}
  }{
    \Matches{p_1(p_2)}{e_1(e_2)}
  } \qquad
  \inferrule[M-Add]{
    \Matches{p_1}{e_1} \\
    \Matches{p_2}{e_2}
  }{
    \Matches{p_1 + p_2}{e_1 + e_2}
  }
\end{mathpar}

\subsubsection{Stripping.} The strip operation erases all filter and
residue wrappers from an expression. It is used by \textsc{M-Lam} and
\textsc{M-Fix} to compare expression bodies modulo instrumentation.

\jbox{\(\Strip{e} = {e'}\)} Strip filter and residue wrappers from expression \(e\).
\[
  \begin{aligned}
    \Strip{x} &= x \\
    \Strip{\Lam{x}{e}} &= \Lam{x}{\Strip{e}} \\
    \Strip{(e_1(e_2))} &= (\Strip{e_1})(\Strip{e_2}) \\
    \Strip{\Nat{n}} &= \Nat{n} \\
    \Strip{(e_1 + e_2)} &= (\Strip{e_1}) + (\Strip{e_2}) \\
    \Strip{(\Filter{f}{e})} &= \Strip{e} \\
    \Strip{\Residue{a}{g}{l}{e}} &= \Strip{e} \\
    \Strip{\Fix{x}{e}} &= \Fix{x}{\Strip{e}}
  \end{aligned}
\]

\subsubsection{Decay.} After a step is taken, one-shot residues
(\(\GasOne\)) are removed while persistent residues (\(\GasAll\)) are
preserved for future steps:

\jbox{\(\Decay{\mathcal{E}} = \mathcal{E}'\)} Evaluation context \(\mathcal{E}\) decays to context \(\mathcal{E}'\).
\[
  \begin{aligned}
    \Decay{x} &= x \\
    \Decay{\Lam{x}{e}} &= \Lam{x}{\Decay{e}} \\
    \Decay{e_1(e_2)} &= (\Decay{e_1})(\Decay{e_2}) \\
    \Decay{\Nat{n}} &= \Nat{n} \\
    \Decay{e_1 + e_2} &= (\Decay{e_1}) + (\Decay{e_2}) \\
    \Decay{\Filter{(p, a, g)}{e}} &= \Filter{(p, a, g)}{\Decay{e}} \\
    \Decay{\Residue{a}{\GasOne}{l}{e}} &= \Decay{e} \\
    \Decay{\Residue{a}{\GasAll}{l}{e}} &= \Residue{a}{\GasAll}{l}{\Decay{e}}
  \end{aligned}
\]

\subsubsection{Action selection.} The analysis judgment walks outward
through the evaluation context, comparing residue priorities to
determine whether to step or skip. When a residue has higher priority
than the current action, it takes over. Otherwise, the existing action
is kept:

\jbox{\(\Analyze{(a, l)}{\mathcal{E}}{a'}\)} Under action \((a, l)\), context \(\mathcal{E}\) yields action \(a'\).
\begin{mathpar}
  \inferrule[A-Var]{
  }{
    \Analyze{(a, l)}{\circ}{a}
  } \\
  \inferrule[A-Ap-L]{
    \Analyze{(a, l)}{\mathcal{E}_1}{a'}
  }{
    \Analyze{(a, l)}{\mathcal{E}_1(e)}{a'}
  } \qquad
  \inferrule[A-Ap-R]{
    \Analyze{(a, l)}{\mathcal{E}_2}{a'}
  }{
    \Analyze{(a, l)}{e_1(\mathcal{E}_2)}{a'}
  } \\
  \inferrule[A-Add-L]{
    \Analyze{(a, l)}{\mathcal{E}_1}{a'}
  }{
    \Analyze{(a, l)}{\mathcal{E}_1 + e}{a'}
  } \qquad
  \inferrule[A-Add-R]{
    \Analyze{(a, l)}{\mathcal{E}_2}{a'}
  }{
    \Analyze{(a, l)}{e_1 + \mathcal{E}_2}{a'}
  } \\
  \inferrule[A-Filter]{
    \Analyze{(a, l)}{\mathcal{E}}{a'}
  }{
    \Analyze{(a, l)}{\Filter{f}{\mathcal{E}}}{a'}
  } \\
  \inferrule[A-Residue-Old]{
    l \le l_0 \\
    \Analyze{(a_0, l_0)}{\mathcal{E}}{a'}
  }{
    \Analyze{(a_0, l_0)}{\Residue{a}{g}{l}{\mathcal{E}}}{a'}
  } \qquad
  \inferrule[A-Residue-New]{
    l > l_0 \\
    \Analyze{(a, l)}{\mathcal{E}}{a'}
  }{
    \Analyze{(a_0, l_0)}{\Residue{a}{g}{l}{\mathcal{E}}}{a'}
  } \\
\end{mathpar}

\subsubsection{Decomposition.} The decomposition judgment splits an
expression \(e\) into an evaluation context \(\mathcal{E}\) and a
redex \(e_0\) such that plugging \(e_0\) into the hole of
\(\mathcal{E}\) recovers \(e\). Decomposition is non-deterministic:
when an expression has multiple redexes (e.g., both operands of an
addition are reducible), there is a choice of which to expose. The
stepper offers this choice to the user when a step is shown, and
picks one deterministically (leftmost, in practice) when stepping
silently. Filter and residue wrappers are traversed transparently
during decomposition. There is one critical exception: a filter or
residue wrapping a \emph{value} is itself a redex
(\textsc{D-Filter-E}, \textsc{D-Residue-E} in
Appendix~\ref{sec:standard_rules}). This is how the
\textsc{T-Filter} and \textsc{T-Residue} transitions above ever
fire: once a sub-expression has finished evaluating to a value, its
surrounding filter and residue wrappers become reducible and are
stripped one at a time. Without this case, residues would
accumulate forever around finalized values.

\jbox{\(\Decompose{e}{\mathcal{E}}{e_0}\)} Expression \(e\)
decomposes into context \(\mathcal{E}\) with redex \(e_0\).

\subsubsection{Composition.} Composition is the inverse of
decomposition: it plugs a redex (or the result of an instruction
transition) back into the hole of an evaluation context. The full
rules are given in Appendix~\ref{sec:standard_rules}.

\jbox{\(\Compose{e}{\mathcal{E}}{e_0}\)} Expression \(e\) is the
result of plugging \(e_0\) into the hole of \(\mathcal{E}\).

\subsubsection{Instruction transitions.} An instruction transition
reduces a single redex in isolation. Function application
substitutes the argument value into the body. Addition reduces to
the numerical sum. Fixpoint unrolls by substituting itself for its
bound variable. Filter and residue wrappers that surround a value
are removed: they have served their purpose once the surrounded
expression has finished evaluating. Substitution \(\FSubst{v}{x}{e}\)
is the standard capture-avoiding operation, defined in
Appendix~\ref{sec:standard_rules}.

\jbox{\(\Transition{e}{e'}\)} Expression \(e\) takes an instruction transition to \(e'\).
\begin{mathpar}
  \inferrule[T-Ap]{
    \Value{e_2}
  }{
    \Transition{(\Lam{x}{e_1})(e_2)}{\FSubst{e_2}{x}{e_1}}
  } \qquad
  \inferrule[T-Add]{
    \Value{n_1} \\
    \Value{n_2} \\
    n_1 + n_2 = n
  }{
    \Transition{\Nat{n_1} + \Nat{n_2}}{\Nat{n}}
  } \\
  \inferrule[T-Fix]{
    \
  }{
    \Transition{\Fix{x}{e}}{\FSubst{\Fix{x}{e}}{x}{e}}
  } \\
  \inferrule[T-Residue]{
    \Value{v}
  }{
    \Transition{\Residue{a}{g}{l}{v}}{v}
  } \qquad
  \inferrule[T-Filter]{
    \Value{v}
  }{
    \Transition{\Filter{f}{v}}{v}
  }
\end{mathpar}

\subsubsection{Optimization.} Instrumentation can stack adjacent
residues around a single redex when several nested filters match it,
and those residues persist into the next step. The optimization pass
collapses adjacent residues to the highest-priority survivor at each
nesting level (the same residue that action selection would have
consulted), so the collapse is observationally invisible. We discuss
the cost of instrumentation and the asymptotic effect of this pass in
Section~\ref{sec:impl-perf}.

\jbox{\(\IsResidue{e}\)} \(e\) is a residue.
\begin{mathpar}
  \inferrule[Is-Residue]{
    \
  }{
    \IsResidue{\Residue{a}{g}{l}{e}}
  }
\end{mathpar}

\jbox{\(\Optimize{e}{e'}\)} \(e\) is optimized to \(e'\)
\begin{mathpar}
  \inferrule[O-Var]{
    \
  }{
    \Optimize{x}{x}
  } \quad
  \inferrule[O-Val]{
    \Value{e}
  }{
    \Optimize{e}{e}
  } \\
  \inferrule[O-Ap]{
    \Optimize{e_l}{e_l'} \\
    \Optimize{e_r}{e_r'}
  }{
    \Optimize{e_l(e_r)}{e_l'(e_r')}
  } \quad
  \inferrule[O-Add]{
    \Optimize{e_l}{e_l'} \\
    \Optimize{e_r}{e_r'}
  }{
    \Optimize{e_l + e_r}{e_l' + e_r'}
  } \\
  \inferrule[O-Filter]{
    \Optimize{e}{e'}
  }{
    \Optimize{\Filter{f}{e}}{\Filter{f}{e'}}
  } \\
  \inferrule[O-Residue-Inner]{
    l_i > l_o \\
    \Optimize{\Residue{a_i}{g_i}{l_i}{e}}{e'}
  }{
    \Optimize{\Residue{a_o}{g_o}{l_o}{\Residue{a_i}{g_i}{l_i}{e}}}{e'}
  } \quad
  \inferrule[O-Residue-Outer]{
    l_i \leq l_o \\
    \Optimize{\Residue{a_o}{g_o}{l_o}{e}}{e'}
  }{
    \Optimize{\Residue{a_o}{g_o}{l_o}{\Residue{a_i}{g_i}{l_i}{e}}}{e'}
  } \quad
  \inferrule[O-Residue-Other]{
    \neg{(\IsResidue{e})} \\
    \Optimize{e}{e'}
  }{
    \Optimize{\Residue{a}{g}{l}{e}}{\Residue{a}{g}{l}{e'}}
  } \\
  \inferrule[O-Fix]{
    \
  }{
    \Optimize{\Fix{x}{e}}{\Fix{x}{e}}
  }
\end{mathpar}

\subsubsection{Stepping judgment.} With all the pieces in place, the
top-level stepping judgment ties the pipeline together.
\textsc{S-Step} fires when action selection yields \(\ActStep\) and
the redex is not a filter or residue elimination (a condition we
formalize with the \(\Strippable{}\) judgment below); the step is
then shown to the user. The two \textsc{S-Skip} rules cover the
silent cases (one for when action selection yields \(\ActSkip\),
one for when the redex is strippable), each performing the step
silently and continuing stepping recursively.

\jbox{\(\Strippable{e}\)} Expression \(e\) is a filter or residue wrapper.
\begin{mathpar}
  \inferrule[F-Filter]{
  }{
    \Strippable{(\Filter{f}{e})}
  } \qquad
  \inferrule[F-Residue]{
  }{
    \Strippable{\Residue{a}{g}{l}{e}}
  }
\end{mathpar}

\jbox{\(\Step{e}{e'}{n}\)} Expression \(e\) steps to \(e'\) in \(n\) steps.
\begin{mathpar}
  \inferrule[S-Step]{
    \InstructPAGL{\PatExpr}{\ActStep}{\GasOne}{0}{e}{e_i} \\
    \Decompose{e_i}{\mathcal{E}_0}{e_0} \\
    \Analyze{(\ActStep, 0)}{\mathcal{E}_0}{\ActStep} \\
    \neg{} (\Strippable{e_0}) \\
    \Transition{e_0}{e_t} \\
    \Compose{e_1}{(\Decay{\mathcal{E}_0})}{e_t}
  }{
    \Step{e}{e_1}{1}
  } \\
  \inferrule[S-Skip-Action]{
    \InstructPAGL{\PatExpr}{\ActStep}{\GasOne}{0}{e}{e_i} \\
    \Decompose{e_i}{\mathcal{E}_0}{e_0} \\
    \Analyze{(\ActStep, 0)}{\mathcal{E}_0}{\ActSkip} \\
    \Transition{e_0}{e_t} \\
    \Compose{e_1}{(\Decay{\mathcal{E}_0})}{e_t} \\
    \Step{e_1}{e_2}{n}
  }{
    \Step{e}{e_2}{n + 1}
  } \\
  \inferrule[S-Skip-Strippable]{
    \InstructPAGL{\PatExpr}{\ActStep}{\GasOne}{0}{e}{e_i} \\
    \Decompose{e_i}{\mathcal{E}_0}{e_0} \\
    \Strippable{e_0} \\
    \Transition{e_0}{e_t} \\
    \Compose{e_1}{(\Decay{\mathcal{E}_0})}{e_t} \\
    \Step{e_1}{e_2}{n}
  }{
    \Step{e}{e_2}{n + 1}
  } \\
  \inferrule[S-Value]{
    \Value{v}
  }{
    \Step{v}{v}{0}
  }
\end{mathpar}

\subsection{Static Semantics}\label{sec:statics}

For the metatheory below, we equip the calculus with a simple type
system. The type grammar adds natural numbers and arrows. The rest
of the syntactic categories from \autoref{fig:filter-syntax} are
unchanged.

\begin{figure}[h]
  \begin{equation*}
    \begin{array}{rcl}
      \DefTyp \tau &\Coloneqq& \Natural \mid \Arrow{\tau}{\tau}
    \end{array}
  \end{equation*}
  \caption{Type grammar.}
  \label{fig:typed-filter-syntax}
\end{figure}

The expression typing judgment \(\TypeEntails{\Gamma}{e}{\tau}\)
threads a typing context \(\Gamma\) of variable bindings through the
expression in the standard way. Filter and residue wrappers are
transparent to typing: they preserve the body's type. Patterns are
typed by a separate but analogous judgment so that we can require
patterns to be closed and well-typed in their declaration
context. The two pattern wildcards \(\PatExpr\) and \(\PatValue\)
have any type, modeling the fact that they match any subject.

\jbox{\(\TypeEntails{\Gamma}{e}{\tau}\)} Under typing context \(\Gamma\), expression \(e\) has type \(\tau\).
\begin{mathpar}
  \inferrule[TE-Var]{
    \TypeContains{\Gamma}{x}{\tau}
  }{
    \TypeEntails{\Gamma}{x}{\tau}
  } \qquad
  \inferrule[TE-Lam]{
    \TypeEntails{\TypeExtends{\Gamma}{x}{\tau_x}}{e}{\tau_e}
  }{
    \TypeEntails{\Gamma}{\Lam{x}{e}}{\Arrow{\tau_x}{\tau_e}}
  } \qquad
  \inferrule[TE-Ap]{
    \TypeEntails{\Gamma}{e_1}{\Arrow{\tau_x}{\tau_e}} \\
    \TypeEntails{\Gamma}{e_2}{\tau_x}
  }{
    \TypeEntails{\Gamma}{e_1(e_2)}{\tau_e}
  } \\
  \inferrule[TE-Nat]{
    \
  }{
    \TypeEntails{\Gamma}{\Nat{n}}{\Natural}
  } \qquad
  \inferrule[TE-Add]{
    \TypeEntails{\Gamma}{e_1}{\Natural} \\
    \TypeEntails{\Gamma}{e_2}{\Natural}
  }{
    \TypeEntails{\Gamma}{e_1 + e_2}{\Natural}
  } \qquad
  \inferrule[TE-Fix]{
    \TypeEntails{\TypeExtends{\Gamma}{x}{\tau}}{e}{\tau}
  }{
    \TypeEntails{\Gamma}{\Fix{x}{e}}{\tau}
  } \\
  \inferrule[TE-Filter]{
    \TypeEntails{\Gamma}{p}{\tau_p} \\
    \TypeEntails{\Gamma}{e}{\tau_e}
  }{
    \TypeEntails{\Gamma}{\Filter{(p, a, g)}{e}}{\tau_e}
  } \qquad
  \inferrule[TE-Residue]{
    \TypeEntails{\Gamma}{e}{\tau}
  }{
    \TypeEntails{\Gamma}{\Residue{a}{g}{l}{e}}{\tau}
  }
\end{mathpar}

\jbox{\(\TypeEntails{\Gamma}{p}{\tau}\)} Under typing context \(\Gamma\), pattern \(p\) has type \(\tau\).
\begin{mathpar}
  \inferrule[TP-Exp]{
    \
  }{
    \TypeEntails{\Gamma}{\PatExpr}{\tau}
  } \qquad
  \inferrule[TP-Val]{
    \
  }{
    \TypeEntails{\Gamma}{\PatValue}{\tau}
  } \qquad
  \inferrule[TP-Var]{
    \TypeContains{\Gamma}{x}{\tau}
  }{
    \TypeEntails{\Gamma}{x}{\tau}
  } \\
  \inferrule[TP-Lam]{
    \TypeEntails{\TypeExtends{\Gamma}{x}{\tau_x}}{p}{\tau_p}
  }{
    \TypeEntails{\Gamma}{\Lam{x}{p}}{\Arrow{\tau_x}{\tau_p}}
  } \qquad
  \inferrule[TP-Ap]{
    \TypeEntails{\Gamma}{p_1}{\Arrow{\tau_x}{\tau_p}} \\
    \TypeEntails{\Gamma}{p_2}{\tau_x}
  }{
    \TypeEntails{\Gamma}{p_1(p_2)}{\tau_p}
  } \\
  \inferrule[TP-Nat]{
    \
  }{
    \TypeEntails{\Gamma}{\Nat{n}}{\Natural}
  } \qquad
  \inferrule[TP-Add]{
    \TypeEntails{\Gamma}{p_1}{\Natural} \\
    \TypeEntails{\Gamma}{p_2}{\Natural}
  }{
    \TypeEntails{\Gamma}{p_1 + p_2}{\Natural}
  } \qquad
  \inferrule[TP-Fix]{
    \TypeEntails{\TypeExtends{\Gamma}{x}{\tau}}{p}{\tau}
  }{
    \TypeEntails{\Gamma}{\Fix{x}{p}}{\tau}
  }
\end{mathpar}

\subsection{Properties}

Because the filtered stepper calculus layers filter and residue
annotations on top of a standard reduction calculus, we want to
confirm that these additions are \emph{conservative}: they should
not break the guarantees of the underlying language. Preservation
(Theorem~\ref{thm:preservation}) ensures that every intermediate
state the stepper shows the user has the same type as the original
expression, so the displayed evaluation remains coherent. Progress
(Theorem~\ref{thm:progress}) ensures that the stepper does not
introduce new stuck states for well-typed programs. Simulation
(Theorem~\ref{thm:simulation}) ensures that the filter machinery
only controls \emph{which} steps are shown, not which reductions
are possible, so the underlying evaluation behavior matches the
base lambda calculus.

\begin{theorem}[Preservation]\label{thm:preservation}
  If \(\TypeEntails{\Gamma}{e}{\tau}\) and \(\Step{e}{e'}{n}\), then \(\TypeEntails{\Gamma}{e'}{\tau}\).
\end{theorem}

\begin{theorem}[Progress]\label{thm:progress}
  If \(\TypeEntails{\varnothing}{e}{\tau}\), then either \(e\) is a value or there exists an \(e'\) such that \(\JustStep{e}{e'}{1}\).
\end{theorem}

Progress is stated over the underlying single-step judgment
\(\JustStep{e}{e'}{1}\), which is the standard formulation. It
follows that a well-typed program either reaches a value through
finitely many filtered steps or diverges. The stepper does not
introduce new stuck states.

A natural stronger statement, ``every well-typed expression either
is a value or takes a visible filtered step,'' is \emph{not}
provable in general. Under a filter that hides every step (e.g.,
\code{debug eval(\$e)}) applied to a divergent program, the stepper
takes no user-visible action, and there is no finite filtered
derivation to construct. What does hold is \emph{single-iteration
progress} (Theorem~\ref{thm:single-iteration}): every well-typed
non-value admits exactly one full iteration of the stepping
pipeline: instrumentation, decomposition, action selection,
transition, and composition. This is the precise guarantee that
``the step button always does something'' even when no visible
step results.

\begin{theorem}[Single-Iteration Progress]\label{thm:single-iteration}
  If \(\TypeEntails{\varnothing}{e}{\tau}\), then either \(e\) is a
  value, or there exist an instrumented expression \(e_i\), a
  context \(\mathcal{E}\), a redex \(e_0\), an action \(a'\), a
  transitioned redex \(e_0'\), and a result \(e'\) such that
  \(\InstructPAGL{\PatExpr}{\ActStep}{\GasOne}{0}{e}{e_i}\),
  \(\Decompose{e_i}{\mathcal{E}}{e_0}\),
  \(\Analyze{(\ActStep, 0)}{\mathcal{E}}{a'}\),
  \(\Transition{e_0}{e_0'}\), and
  \(\Compose{e'}{(\Decay{\mathcal{E}})}{e_0'}\).
\end{theorem}

To state the simulation theorem, we define an unfiltered stepping
judgment that ignores all filter and residue annotations, together
with its reflexive--transitive closure
\(\JustStepStar{e}{e'}\).

\jbox{\(\JustStep{e}{e'}{1}\)} Expression \(e\) takes one unfiltered step to \(e'\).
\begin{mathpar}
  \inferrule[J-Step]{
    \Decompose{e}{\mathcal{E}_0}{e_0} \\
    \Transition{e_0}{e_0'} \\
    \Compose{e'}{\mathcal{E}_0}{e_0'}
  }{
    \JustStep{e}{e'}{1}
  }
\end{mathpar}

\jbox{\(\JustStepStar{e}{e'}\)} Expression \(e\) takes zero or more unfiltered steps to \(e'\).
\begin{mathpar}
  \inferrule[J-Refl]{ }{
    \JustStepStar{e}{e}
  } \qquad
  \inferrule[J-Trans]{
    \JustStep{e}{e_1}{1} \\
    \JustStepStar{e_1}{e'}
  }{
    \JustStepStar{e}{e'}
  }
\end{mathpar}

\begin{theorem}[Simulation]\label{thm:simulation}
  If \(\Step{e}{e'}{n}\), then \(\JustStepStar{\Strip{e}}{\Strip{e'}}\).
\end{theorem}

Each silent transition the stepper takes corresponds to one
unfiltered step of the stripped program, and each user-visible
transition corresponds to a single unfiltered step as well, so an
\(n\)-step filtered derivation between \(e\) and \(e'\) is simulated
by exactly \(n\) unfiltered steps between \(\Strip{e}\) and
\(\Strip{e'}\). We do not record this count in the statement, since
the corollary we care about, that the stepper does not invent
reductions, follows from the existence of any such derivation.

\subsection{Agda Mechanization}

We mechanize the dynamic and static semantics of the calculus,
together with proofs of preservation, progress, single-iteration
progress, and simulation as stated above. We use de Bruijn indices
to implement \(\alpha\)-equivalence during the filter-matching
process. To mitigate the complexity of de Bruijn indices, we fuse
the operations of substitution and function application into a
single function, eliminating many structurally possible but invalid
states (for example, shifting down a variable with index 0).
Additionally, we use well-founded recursion to prove termination of
the instrumentation and optimization passes.


\section{Implementation}\label{sec:implementation}


We describe two implementations of the Filtered Stepper Calculus: a
compact reference implementation in OCaml that directly mirrors the
formal rules, and the full integration into the Hazel live programming
environment.

\subsection{Reference Implementation}\label{sec:impl-mini}

To validate the formal semantics presented in Section~\ref{sec:filter},
we built a standalone reference stepper in OCaml. It implements a
substitution-based small-step evaluator for the simply typed lambda
calculus extended with a fixpoint operator and the full filter calculus.
The core is roughly 560 lines of OCaml, structured as a direct
transliteration of the formal judgments: expressions and patterns are
algebraic data types, actions and gas are simple sums, and filter and
residue expressions are constructors of the expression type.

A single stepping function orchestrates the pipeline described in
Section~\ref{sec:dynamics}, calling instrumentation, optimization,
decomposition, action selection, transition, decay, and composition
in sequence. If action selection yields \(\ActSkip\), or if the redex
is a filter or residue elimination (the \(\Strippable{}\) case), the
stepper performs the transition silently and recurses. Otherwise it
returns the list of available redexes to the caller.

Because the reference implementation uses substitution-based evaluation
and named variables, it corresponds closely to the on-paper calculus and
is easy to audit against the formal rules. We test the stepper against
known results (e.g., verifying that \texttt{fac(5)} evaluates to 120).
On the common subset of the two calculi (lambda calculus with addition,
fixpoints, and filters), the reference implementation also serves as a
differential test oracle for the Hazel integration; any divergence
between the two indicates a bug in the production stepper. Hazel
features beyond that subset---pattern matching, sum and product types,
type ascriptions, typed holes---are covered by separate hand-written
unit tests against expected trace outputs.

\subsection{Hazel Integration}\label{sec:impl-hazel}

Hazel internally uses a big-step, environment-based evaluator for
performance: it avoids redundant substitutions and enables tail-call
optimization. However, because the course teaches substitution-based
equational semantics, the stepper must present traces as if
substitution were performed eagerly.

To keep the two evaluation styles consistent, the big-step evaluator
and the small-step stepper share a common transition relation: each
redex's reduction case is defined in one place, and the two
interpreters differ only in their driver loop (recursive
descent for the big-step evaluator versus single-redex extraction
for the stepper). The stepper works over Hazel's internal expression
representation, which includes closures (pairs of an expression and
an environment). During decomposition, closures are traversed like
ordinary expressions, with the environment carried in the evaluation
context. When the stepper displays a step to the user, a
post-processing pass replaces all bound variables with their
environment bindings, producing output that looks substitution-based.

The filter system in Hazel mirrors the reference implementation but
operates over Hazel's richer expression language, which includes
pattern matching, sum and product types, type ascriptions, and typed
holes. Filter matching uses a semantic equality check with special
accommodations: fixpoint wrappers and type ascriptions are ignored
during comparison.

Beyond user-written filters, Hazel exposes UI toggles that classify
certain step kinds (fixpoint unrolling, type-ascription elimination,
pattern-match redex selection) as skipped-by-default for the trace,
while a few others (closure-wrapping, residue bookkeeping) are
unconditionally hidden because they correspond to internal machinery
rather than user-visible reductions.


\subsection{Performance}\label{sec:impl-perf}

We have already discussed the cost of instrumentation in
Section~\ref{sec:dynamics}. The instrumentation rules walk the whole
expression, and at each node test the active filter's pattern against that
node---the I-Y/I-N split. With $k$ filters in scope and an expression of size
$n$, this is $O(n \cdot k \cdot |p|)$ work per step, where $|p|$ is the largest
pattern. The $k$ factor is unavoidable, since a deeply nested filter must still
be considered at every node it covers.

The structural problem is what happens when several nested filters
match the same redex. Two effects compound. First, within a single
instrumentation pass, the recursive structure of \textsc{I-Filter}
re-walks already-instrumented sub-expressions and can stack residues
exponentially in the nesting depth (Section~\ref{sec:dynamics}).
Second, those residues are not transient: they persist as part of
the expression into the next step, where re-instrumentation walks
over them and may add more on top. Without intervention, residue
chains grow with the \emph{history} of stepping as well as with
filter nesting depth, so the cost of decomposition and action
selection drifts away from $O(n)$ and keeps climbing.

The optimization pass cuts this back. \textsc{O-Residue-Inner} and
\textsc{O-Residue-Outer} collapse each maximal chain of adjacent
residues to its highest-priority survivor, so residue depth per chain
stays at $O(1)$ and the total residue count stays at $O(n)$ regardless
of how many filters are in scope. Decomposition and action selection
then work on an expression tree of size $O(n)$ rather than
$O(n \cdot k)$. The $|p|$ factor in instrumentation remains the
asymptotic bottleneck per step. This is unsurprising, since pattern
matching is the work the user is actually paying for.


\section{Evaluation}\label{sec:evaluation}


To assess how the filtered stepper supports instructors and students
learning functional programming, we deployed it in an upper-level
university programming languages course. We first report on the
instructor's experience using the stepper in lecture, then on student
usage and survey responses from homework assignments.

\subsection{Lecture Integration}

Before the filtered stepper was integrated into Hazel, the instructor
of the course presented small-step reductions in lecture by hand,
working through example traces on the whiteboard. This required significant
time writing out reductions that were mostly similar between steps.
Since the stepper was deployed, the instructor has
used it to run these example evaluations live in lecture instead.
This shift had three practical effects. First, walking through an
evaluation became substantially less tedious. Filter annotations
let the instructor collapse routine reductions and surface only the
steps that matter for the lesson at hand, without having to commit
to those choices in slide content ahead of time. Second, when a
student asked a follow-up question, the instructor was able to edit the
example and rerun the stepper in real time. With hand-worked paper
traces, exploring a ``what if we changed this?'' would have stalled
the lecture. Third, the
stepper served as a running demonstration of Hazel itself: the same
environment the students were using for their assignments was on
display in lecture, with its hole-driven editor, live evaluation,
and stepper visible side by side.

\subsection{Assignment Integration}

The assignments we evaluated our stepper on usually consist of several tasks, each of which asks students to implement or complete a function to pass given unit tests. To help students focus on their task, we provide prelude Hazel definitions that they can use directly in their assignments. We do not want students to inspect the evaluation of the prelude, as this would break the narrative that the prelude already exists in the environment. We therefore insert a \lstinline[language=hazel]{debug hide($e)} at the very top of the program, and a \lstinline[language=hazel]{debug stop($e)} before the user program. Whenever the user turns the stepper on, it skips evaluation of the prelude and stops at the entrance of the user program.

\begin{figure}[ht]
  \centering
  \begin{minipage}{.40\linewidth}
    \begin{subfigure}{\linewidth}
      \includegraphics[width=\textwidth]{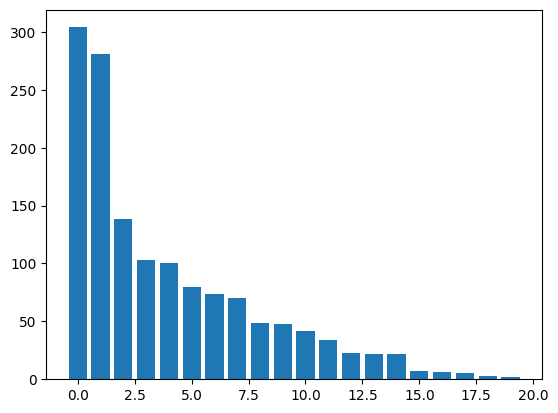}
      \caption{Winter 2024, sample size 33}
      \label{fig:eval-num-steps-w24}
    \end{subfigure}
  \end{minipage}
  \quad
  \begin{minipage}{.40\linewidth}
    \begin{subfigure}{\linewidth}
      \includegraphics[width=\textwidth]{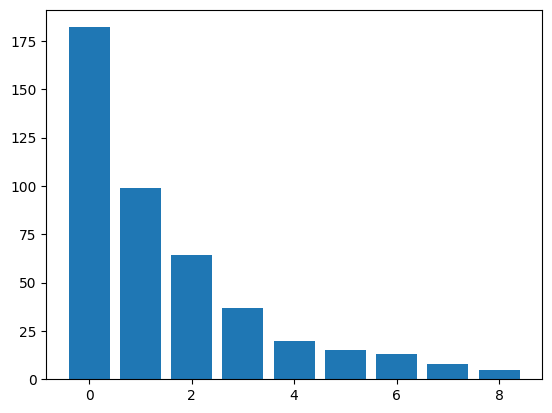}
      \caption{Fall 2024, sample size 20}
      \label{fig:eval-num-steps-f24}
    \end{subfigure}
  \end{minipage}
  \caption{Number of steps used by students}
  \label{fig:eval-num-steps}
\end{figure}

We collected anonymous data from students during one of their
assignments. The number of steps used by students is shown in
Figure~\ref{fig:eval-num-steps}.

Students were never explicitly instructed in how to use the stepper,
and were not asked to use it on the homework. Even so, nearly half
of them saw the instructor demonstrate it in lecture and chose to
use it themselves on the assignment.

The number of steps used by students roughly conforms to an
exponential distribution. A small portion of students use the
stepper extensively.

To complement the usage data, we administered an anonymous end-of-term
survey to the same cohort. Of 25 respondents, 20 reported using the
stepper and 12 of those adopted at least one of the
\lstinline[language=hazel]{hide}, \lstinline[language=hazel]{stop},
\lstinline[language=hazel]{eval}, or \lstinline[language=hazel]{step}
filter keywords. The Likert responses are summarized in
Table~\ref{tab:eval-survey-likert}. We classify usage using the direct
usage question and summarize Likert responses only for respondents who
reported using the stepper.

\begin{table}[ht]
  \centering
  \caption{Self-reported experience with the stepper among users, Strongly/Somewhat collapsed ($n=20$).}
  \label{tab:eval-survey-likert}
  \begin{tabular}{lccc}
    \toprule
    Statement (paraphrased) & Agree & Neither & Disagree \\
    \midrule
    Difficult to read                       & 5  & 4 & 11 \\
    Surprised by green highlighting         & 9  & 7 & 4  \\
    Helped identify or correct bugs         & 12 & 5 & 3  \\
    Helped understand Hazel's features      & 13 & 6 & 1  \\
    Long and tedious to interact with       & 7  & 5 & 8  \\
    \bottomrule
  \end{tabular}
\end{table}

Three observations triangulate with the usage data. First, the
strongest positive response (13 vs.\ 1) is for \emph{understanding}
language features rather than bug-finding (12 vs.\ 3), and the
free-text ``what did you use it for?'' question agrees: ``understanding
how Hazel evaluates'' is mentioned slightly more often than
``debugging,'' consistent with the stepper's substitution-based,
source-level design. Second, readability is not the issue (11 vs.\ 5),
but ``long and tedious'' splits evenly (7 vs.\ 8) and tracks the
long-tailed shape of Figure~\ref{fig:eval-num-steps}; the
12-of-20 adoption rate of the filter keywords suggests students
reach for filters as friction reduction, as designed. Third, the
green-highlighting question drew nearly half the users to agree
(9 of 20), indicating that the lexical scoping of filters is not
self-evident in the current source-text-only interface, motivation
for the graphical filter panel discussed as future work.

The survey is small ($n=25$), single-cohort, voluntary, and
self-report. One respondent reported uniformly negative experiences
attributed to bugs in the deployed prototype.


\section{Related Work}\label{sec:related}

\paragraph{Algebraic steppers.}
Algebraic steppers for functional languages have long served pedagogy. The
DrScheme/Racket algebraic stepper presents small-step reductions that mirror
operational semantics, with formal correctness arguments for its
traces~\cite{clements2001stepper}. A parallel rewriting-based tradition runs
through the HIPE/WinHIPE line, in which user programs are evaluated by an
exposed rewriting system and the resulting traces are animated for classroom
use~\cite{urquiza2007winhipe}. Cong and Asai showed how delimited continuations
can implement steppers for ML-like languages~\cite{cong2016stepper}, a line
continued in steppers for practical OCaml fragments that handle effectful
constructs such as exceptions and printing
primitives~\cite{furukawa2019stepping}, and most recently extended to simple
modules with hierarchical variable references~\cite{asai2025modules}. Closest
to our motivation, Olmer, Heeren, and Jeuring's Haskell tutoring environment
addresses trace verbosity by letting students switch between innermost and
outermost evaluation orders to control which reductions are
shown~\cite{olmer2014evaluating}; the surrounding Ask-Elle tutor integrates
this with automated feedback on student-written Haskell
programs~\cite{gerdes2017askelle}. Our work keeps the source-level,
semantics-directed view of evaluation common to all of these systems, but
gives users a finer-grained control: rather than picking a strategy globally,
users embed lexically scoped patterns that select which reductions appear in
the trace on a redex-by-redex basis.

\paragraph{Programmer-directed focus.}
Bertot's occurrence-based debugger specifications and Bernstein and Stark's
focusing debugger formalize how users can direct debugging attention to
particular source-level
occurrences~\cite{bertot1991occurrences,bernstein1995focusing}. Algorithmic
debugging, going back to Shapiro~\cite{shapiro1983algorithmic} and continuing
through modern Haskell instantiations such as Buddha and
Hoed~\cite{pope2004buddha,faddegon2016hoed}, achieves user-directed focus
through oracle questioning: the programmer's yes/no answers about
subcomputation correctness successively narrow the debugger's attention to
the buggy subtree. Closely related, Perera et al.\ recast a functional
evaluation trace as an \emph{explanation} of an observed output, using
program slicing to extract the minimal subprogram that produced
it~\cite{perera2012explain}. Interrogative debugging via the
Whyline takes a complementary approach, generating a menu of ``why did'' and
``why didn't'' questions that the user selects to highlight relevant runtime
events~\cite{ko2004whyline}. Expositor exposes execution traces as first-class
values that programmers can script with \texttt{map} and \texttt{filter} to
direct focus~\cite{khoo2013expositor}. On the imperative side, Boothe's
bidirectional debugger gave programmers an event-counter-based language for
navigating execution forwards and backwards along chosen event
classes~\cite{boothe2000bidirectional}. Hazel filters extend this lineage with
redex-structural patterns that compose through lexical scope and integrate
with an algebraic stepper.

\paragraph{Lexically scoped source annotations.}
The methodological pattern of attaching lightweight, lexically scoped
annotations to source code to control a dynamic observation has a direct
ancestor in profiling. Sansom and Peyton Jones introduced \emph{cost centres}
for higher-order functional languages: programmer-written annotations,
attached to source regions, that attribute runtime costs to named scopes,
with a formal cost-attribution semantics that is invariant under
evaluation~\cite{sansom1997profiling}. Filter annotations share this
methodological shape. A lexically scoped construct in the source does not
affect the result of evaluation, but governs how a dynamic event (cost
attribution there, step visibility here) is attributed to source regions.

\paragraph{Tracing tools for functional languages.}
Tracing tools record the reduction history of a computation for later
inspection. An early source-level debugger from Standard ML by Tolmach and
Appel used source-to-source instrumentation and first-class continuations to
provide breakpoints and reverse execution without depending on the machine
representation~\cite{tolmach1995debugger}. Nilsson and Fritzson's Freja
debugger introduced declarative debugging for lazy functional languages, using
an evaluation-dependence tree as the basis for oracle
questioning~\cite{nilsson1994algorithmic}. Hood lets programmers insert
\texttt{observe} annotations to capture intermediate values without altering
program semantics~\cite{gill2000hood}, and GHood adds graphical animation of
those observations~\cite{reinke2001ghood}. Hat records an \emph{augmented
redex trail} and provides multiple views over it, including search and
structured querying~\cite{wallace2001hat}. Chitil's source-based trace
explorer builds on Hat to unify algorithmic debugging, stepping, and dynamic
slicing into a single source-driven viewer in which the user navigates the
trace by selecting source-level subterms~\cite{chitil2005source}. The GHCi
debugger brought interactive source-level breakpoints to lazy Haskell via
lightweight bytecode instrumentation~\cite{marlow2007ghci}. More recently,
CSI:~Haskell extends GHC's coverage tooling to record evaluation traces
alongside the call stack, reconstructing the chain of events leading to a
fault in lazy code~\cite{gissurarson2023csi}. Our stepper similarly maintains
a step history, but the trace is \emph{filtered} by user-specified patterns
rather than recorded uniformly.

\paragraph{Pattern languages and term-rewriting strategies.}
Our filter language is also related to program-pattern and
strategy-controlled rewriting systems. SCRUPLE and ASTLOG use structural
patterns to query program trees~\cite{paul1994scruple,crew1997astlog}.
Stratego separates rewrite rules from strategies that control where and how
rules are applied~\cite{visser2001stratego}. Coccinelle's \emph{semantic
patches} generalize Unix-style patches to a structural pattern language over
C, and have driven thousands of automated evolutions in the Linux
kernel~\cite{padioleau2008coccinelle}. Hazel filters use a smaller pattern
language and do not let users change the evaluator. Instead, patterns control
the visibility of steps produced by a fixed algebraic stepper.

\paragraph{Programming education context.}
More broadly, our stepper sits within a long-running line of work on program
visualization and \emph{notional machines} for introductory programming
education. Sorva, Karavirta, and Malmi survey generic visualization systems
and distill the design lessons of that body of work, including the central
tension between fidelity to the underlying semantics and clarity for a
novice~\cite{sorva2013review}. Clements and Krishnamurthi's recent
investigation of pedagogical models for runtime stacks and scope --- using a
block-based programming environment for empirical study --- illustrates that
the question of how best to render evaluation comprehensible to students
remains an active research concern~\cite{clements2022notional}.

\paragraph{Hazel substrate.}
Hazelnut-style structural editing and semantics for incomplete programs make
the AST explicit at all times, which naturally supports tree-pattern matching
and stepping in the presence of holes~\cite{omar2017hazelnut,omar2019live}.
The substrate has since been extended with typed pattern holes that enable
live evaluation in the presence of incomplete match
expressions~\cite{yuan2023peanut}, and with the marked lambda calculus that
gives a total account of type error localization and recovery, providing the
foundation on which modern Hazel is built~\cite{zhao2024marking}. The filter
calculus builds on this substrate to give a precise semantics for scoped,
pattern-based stepping control in a live functional environment.

\section{Discussion and Conclusion}\label{sec:discussion}

We have presented the Filtered Stepper Calculus, a formal framework
for scoped, pattern-based control over which reduction steps are
visible during algebraic stepping. The calculus is backed by
mechanized proofs of preservation, progress, and simulation in Agda,
and is implemented in the Hazel live programming environment, where it
has been deployed in a university programming languages course.

\paragraph{Stepper adoption.}
Our evaluation shows that students adopted the stepper organically:
without being instructed to use it, nearly half of the students who
observed the stepper in lecture chose to use it on their assignments.
This suggests that an algebraic stepper that mirrors the
substitution-based reasoning taught in class fills a genuine need
in the pedagogical workflow.

\paragraph{Filter adoption.}
By contrast, we observed limited student use of the filter system
during assignments. We attribute this to three factors. First, the
current interface requires users to write filter annotations directly
in their program text, which is unfamiliar syntax and adds friction.
Second, inserting a filter currently triggers re-elaboration of the
entire program, introducing noticeable latency. Third, students
working on assignments are focused on correctness rather than trace
exploration, so the cost of learning the filter language may not be
justified in that context.

\paragraph{Lecture versus assignment contexts.}
A natural reading is that filters answer different needs in lecture
versus assignment contexts. In lecture, the instructor prepares
filters ahead of time to direct attention, and the cost of writing
them is amortized over a classroom of viewers. On assignments, a
student writes filters for an audience of one and pays the full cost
themselves, with no immediate payoff in terms of grade or
correctness. The lecture-side adoption suggests filters are most
useful, at least initially, as a presentation tool. Broader student
adoption may depend on lowering the per-use cost (cf. the graphical
panel discussed below).

\paragraph{Limitations.}
The primary contribution of this paper is a theoretical treatment of
pattern-based control over algebraic stepping: a formal calculus,
mechanized metatheory, and a working implementation. The classroom
deployment we report is illustrative rather than a controlled study.
It involves one instructor at one institution, the assignment sample
is small ($n = 33 + 20$), survey responses are voluntary and
self-reported, and we did not measure learning outcomes. Stronger
claims about pedagogical effectiveness would require a controlled
study against an unfiltered stepper and a no-stepper baseline,
ideally across multiple instructors. Our evaluation is intended
only to demonstrate that students and an instructor will pick up the
tool and use it under normal classroom conditions.

\paragraph{Future work.}

The filter-adoption findings point to a clear design direction:
separating the filter interface from the program text. We envision a
graphical filter panel in which users can add, modify, and remove
filters interactively \emph{during stepping}, without editing source
code or triggering re-evaluation. This would lower the barrier to
entry and, perhaps more importantly, make filters useful for
\emph{interactive debugging}: a user could discover what to focus on
while exploring a trace, rather than predicting it ahead of time.
Filter annotations would then function less like up-front program
declarations and more like queries on an ongoing computation.

A second direction is a more expressive filter language. The current
pattern language matches on the syntactic structure of the redex up
to a fixed nesting depth. Two kinds of queries we found ourselves
wanting are not expressible. The first refers to the type or shape
of sub-expressions: ``skip all \texttt{let} reductions whose
right-hand side is a function value'' is a natural request, but our
patterns can only match values uniformly, not values of a particular
type. The second requires recursive structure: ``stop at any
expression composed entirely of literal values and arithmetic
operators, before it begins reducing'' would require something like
a grammar-style production or a recursive predicate over the AST.
Type-based patterns and predicate-based patterns would each
address part of this gap.

A third direction is filter inference. When a user is overwhelmed by
a long trace, the system could observe which steps they advance
through quickly versus dwell on and propose filter annotations
automatically.

Relatedly, instructors could ship pedagogical filter sets with
assignments---a shared library of common filter patterns that
students apply directly. Filters become reusable classroom artifacts
in their own right.

Finally, adapting the filter calculus to richer evaluation
strategies, such as lazy evaluation or concurrent reduction, would
broaden its applicability beyond Hazel.

\paragraph{Conclusion.}
The Filtered Stepper Calculus offers a principled, mechanized answer
to the ``too many steps'' problem in algebraic stepping: let the user
describe, lexically and by structure, which reductions to surface,
and let the calculus handle the rest. The remaining design space
lies largely in user interface---how best to expose the calculus's
expressive power to instructors and students in practice.

\bibliographystyle{splncs04}
\bibliography{references}

\begin{thebibliography}{10}
\providecommand{\url}[1]{\texttt{#1}}
\providecommand{\urlprefix}{URL }
\providecommand{\doi}[1]{https://doi.org/#1}

\bibitem{asai2025modules}
Asai, K., Akiyama, H.: Algebraic stepper for simple modules. In: Proceedings of
  the 2025 ACM SIGPLAN International Workshop on Partial Evaluation and Program
  Manipulation. p. 13–29. PEPM '25, Association for Computing Machinery, New
  York, NY, USA (2025). \doi{10.1145/3704253.3706137}

\bibitem{bernstein1995focusing}
Bernstein, K.L., Stark, E.W.: Operational semantics of a focusing debugger.
  Electronic Notes in Theoretical Computer Science  \textbf{1},  13--31 (1995).
  \doi{10.1016/S1571-0661(04)80002-1},
  \url{https://www.sciencedirect.com/science/article/pii/S1571066104800021},
  mFPS XI, Mathematical Foundations of Programming Semantics, Eleventh Annual
  Conference

\bibitem{bertot1991occurrences}
Bertot, Y.: Occurrences in debugger specifications. In: Proceedings of the ACM
  SIGPLAN 1991 Conference on Programming Language Design and Implementation. p.
  327–337. PLDI '91, Association for Computing Machinery, New York, NY, USA
  (1991). \doi{10.1145/113445.113473}

\bibitem{boothe2000bidirectional}
Boothe, B.: Efficient algorithms for bidirectional debugging. In: Proceedings
  of the {ACM} {SIGPLAN} 2000 Conference on Programming Language Design and
  Implementation ({PLDI} 2000). pp. 299--310. ACM (2000).
  \doi{10.1145/349299.349339}

\bibitem{chitil2005source}
Chitil, O.: Source-based trace exploration. In: Implementation and Application
  of Functional Languages, {IFL} 2004. Lecture Notes in Computer Science,
  vol.~3474, pp. 126--141. Springer (2005). \doi{10.1007/11431664_8}

\bibitem{clements2001stepper}
Clements, J., Flatt, M., Felleisen, M.: Modeling an algebraic stepper. In:
  Sands, D. (ed.) Programming Languages and Systems. pp. 320--334. Springer
  Berlin Heidelberg, Berlin, Heidelberg (2001). \doi{10.1007/3-540-45309-1_21}

\bibitem{clements2022notional}
Clements, J., Krishnamurthi, S.: Towards a notional machine for runtime stacks
  and scope: When stacks don't stack up. In: Proceedings of the 2022 {ACM}
  Conference on International Computing Education Research ({ICER} 2022). pp.
  206--222. ACM (2022). \doi{10.1145/3501385.3543961}

\bibitem{cong2016stepper}
Cong, Y., Asai, K.: Implementing a stepper using delimited continuations. In:
  Davenport, J.H., Ghourabi, F. (eds.) SCSS 2016. 7th International Symposium
  on Symbolic Computation in Software Science. EPiC Series in Computing,
  vol.~39, pp. 42--54. EasyChair (2016). \doi{10.29007/l2wb},
  \url{https://easychair.org/publications/paper/7qlb}

\bibitem{crew1997astlog}
Crew, R.F.: Astlog: a language for examining abstract syntax trees. In:
  Proceedings of the Conference on Domain-Specific Languages on Conference on
  Domain-Specific Languages (DSL), 1997. p.~18. DSL'97, USENIX Association, USA
  (1997). \doi{10.5555/1267950.1267968}

\bibitem{faddegon2016hoed}
Faddegon, M., Chitil, O.: Lightweight computation tree tracing for lazy
  functional languages. In: Proceedings of the 37th ACM SIGPLAN Conference on
  Programming Language Design and Implementation ({PLDI} 2016). ACM (2016).
  \doi{10.1145/2908080.2908104}

\bibitem{furukawa2019stepping}
Furukawa, T., Cong, Y., Asai, K.: Stepping ocaml. Electronic Proceedings in
  Theoretical Computer Science  \textbf{295},  17--34 (2019).
  \doi{10.4204/EPTCS.295.2}

\bibitem{gerdes2017askelle}
Gerdes, A., Heeren, B., Jeuring, J., van Binsbergen, L.T.: {Ask-Elle}: an
  adaptable programming tutor for {Haskell} giving automated feedback.
  International Journal of Artificial Intelligence in Education
  \textbf{27}(1),  65--100 (2017). \doi{10.1007/s40593-015-0080-x}

\bibitem{gill2000hood}
Gill, A.: Debugging haskell by observing intermediate data structures.
  Electronic Notes in Theoretical Computer Science  \textbf{41}(1), ~1 (2001).
  \doi{10.1016/S1571-0661(05)80538-9},
  \url{https://www.sciencedirect.com/science/article/pii/S1571066105805389},
  2000 ACM SIGPLAN Haskell Workshop (Satellite Event of PLI 2000)

\bibitem{gissurarson2023csi}
Gissurarson, M.P., Applis, L.H.: Csi: Haskell - tracing lazy evaluations in a
  functional language. In: Proceedings of the 35th Symposium on Implementation
  and Application of Functional Languages. IFL '23, Association for Computing
  Machinery, New York, NY, USA (2024). \doi{10.1145/3652561.3652562}

\bibitem{khoo2013expositor}
Khoo, Y.P., Foster, J.S., Hicks, M.: Expositor: Scriptable time-travel
  debugging with first-class traces. In: 2013 35th International Conference on
  Software Engineering (ICSE). pp. 352--361 (2013).
  \doi{10.1109/ICSE.2013.6606581}

\bibitem{ko2004whyline}
Ko, A.J., Myers, B.A.: Designing the whyline: a debugging interface for asking
  questions about program behavior. In: Proceedings of the SIGCHI Conference on
  Human Factors in Computing Systems. p. 151–158. CHI '04, Association for
  Computing Machinery, New York, NY, USA (2004). \doi{10.1145/985692.985712}

\bibitem{marlow2007ghci}
Marlow, S., Iborra, J., Pope, B., Gill, A.: A lightweight interactive debugger
  for haskell. In: Proceedings of the ACM SIGPLAN Workshop on Haskell Workshop.
  p. 13–24. Haskell '07, Association for Computing Machinery, New York, NY,
  USA (2007). \doi{10.1145/1291201.1291204}

\bibitem{nilsson1994algorithmic}
Nilsson, H., Fritzson, P.: Algorithmic debugging for lazy functional languages.
  Journal of Functional Programming  \textbf{4}(3),  337--370 (1994).
  \doi{10.1017/S0956796800001088}

\bibitem{olmer2014evaluating}
Olmer, T., Heeren, B., Jeuring, J.: Evaluating {Haskell} expressions in a
  tutoring environment. In: Proceedings 3rd International Workshop on Trends in
  Functional Programming in Education ({TFPIE} 2014). Electronic Proceedings in
  Theoretical Computer Science, vol.~170, pp. 50--66 (2014)

\bibitem{omar2019live}
Omar, C., Voysey, I., Chugh, R., Hammer, M.A.: Live functional programming with
  typed holes. Proc. ACM Program. Lang.  \textbf{3}(POPL) (Jan 2019).
  \doi{10.1145/3290327}

\bibitem{omar2017hazelnut}
Omar, C., Voysey, I., Hilton, M., Aldrich, J., Hammer, M.A.: Hazelnut: a
  bidirectionally typed structure editor calculus. In: Proceedings of the 44th
  ACM SIGPLAN Symposium on Principles of Programming Languages. p. 86–99.
  POPL '17, Association for Computing Machinery, New York, NY, USA (2017).
  \doi{10.1145/3009837.3009900}

\bibitem{padioleau2008coccinelle}
Padioleau, Y., Lawall, J., Hansen, R.R., Muller, G.: Documenting and automating
  collateral evolutions in linux device drivers. In: Proceedings of the 3rd ACM
  SIGOPS/EuroSys European Conference on Computer Systems 2008. p. 247–260.
  Eurosys '08, Association for Computing Machinery, New York, NY, USA (2008).
  \doi{10.1145/1352592.1352618}

\bibitem{paul1994scruple}
Paul, S., Prakash, A.: A framework for source code search using program
  patterns. IEEE Trans. Softw. Eng.  \textbf{20}(6),  463–475 (Jun 1994).
  \doi{10.1109/32.295894}

\bibitem{perera2012explain}
Perera, R., Acar, U.A., Cheney, J., Levy, P.B.: Functional programs that
  explain their work. In: Proceedings of the 17th {ACM} {SIGPLAN} International
  Conference on Functional Programming ({ICFP} 2012). pp. 365--376. ACM (2012).
  \doi{10.1145/2364527.2364579}

\bibitem{pope2004buddha}
Pope, B.: Declarative debugging with buddha. In: Vene, V., Uustalu, T. (eds.)
  Advanced Functional Programming, Lecture Notes in Computer Science,
  vol.~3622, pp. 273--308. Springer (2005). \doi{10.1007/11546382_7}

\bibitem{reinke2001ghood}
Reinke, C.: {GHood}---graphical visualisation and animation of {Haskell} object
  observations. In: Proceedings of the 2001 {ACM} {SIGPLAN} Haskell Workshop.
  Electronic Notes in Theoretical Computer Science, vol.~59. Elsevier (2001)

\bibitem{sansom1997profiling}
Sansom, P.M., {Peyton Jones}, S.L.: Formally based profiling for higher-order
  functional languages. ACM Transactions on Programming Languages and Systems
  \textbf{19}(2),  334--385 (1997). \doi{10.1145/244795.244797}

\bibitem{shapiro1983algorithmic}
Shapiro, E.Y.: Algorithmic Program Debugging. The MIT Press (04 1983).
  \doi{10.7551/mitpress/1192.001.0001}

\bibitem{sorva2013review}
Sorva, J., Karavirta, V., Malmi, L.: A review of generic program visualization
  systems for introductory programming education. ACM Transactions on Computing
  Education  \textbf{13}(4) (2013). \doi{10.1145/2490822}

\bibitem{tolmach1995debugger}
Tolmach, A., Appel, A.W.: A debugger for {Standard ML}. Journal of Functional
  Programming  \textbf{5}(2),  155--200 (Apr 1995).
  \doi{10.1017/S0956796800001313}

\bibitem{urquiza2007winhipe}
Urquiza-Fuentes, J., Vel{\'a}zquez-Iturbide, J.{\'A}.: An evaluation of the
  effortless approach to build algorithm animations with {WinHIPE}. Electronic
  Notes in Theoretical Computer Science  \textbf{178},  3--13 (2007).
  \doi{10.1016/j.entcs.2007.01.038}, proceedings of the 4th International
  Program Visualization Workshop (PVW 2006)

\bibitem{visser2001stratego}
Visser, E.: Stratego: A language for program transformation based on rewriting
  strategies. In: Proceedings of the 12th International Conference on Rewriting
  Techniques and Applications. p. 357–362. RTA '01, Springer-Verlag, Berlin,
  Heidelberg (2001). \doi{10.1007/3-540-45127-7_27}

\bibitem{wallace2001hat}
Wallace, M., Chitil, O., Brehm, T., Runciman, C.: Multiple-view tracing for
  haskell: a new hat. In: Hinze, R. (ed.) 2001 ACM SIGPLAN Haskell Workshop.
  Firenze, Italy (September 2001), \url{https://kar.kent.ac.uk/13566/},
  universiteit Utrecht UU-CS-2001-23. Final proceedings to appear in ENTCS
  59(2).

\bibitem{yuan2023peanut}
Yuan, Y., Guest, S., Griffis, E., Potter, H., Moon, D., Omar, C.: Live pattern
  matching with typed holes. Proc. ACM Program. Lang.  \textbf{7}(OOPSLA1) (Apr
  2023). \doi{10.1145/3586048}

\bibitem{zhao2024marking}
Zhao, E., Maroof, R., Dukkipati, A., Blinn, A., Pan, Z., Omar, C.: Total type
  error localization and recovery with holes. Proc. ACM Program. Lang.
  \textbf{8}(POPL) (Jan 2024). \doi{10.1145/3632910}

\end{thebibliography}

\appendix
\section{Standard Rules}\label{sec:standard_rules}

This appendix collects the standard meta-functions and rules of the
filtered stepper calculus (substitution, decomposition, and
composition) deferred from \autoref{sec:filter}. The syntax of
evaluation contexts and the value judgment are given in
\autoref{fig:filter-syntax} and the accompanying rules in the main
body.

\fbox{\([v / x] e = e'\)} \(e'\) can be obtained by substitution of \(v\) for
\(x\) in expression \(e\).
\[
  \begin{aligned}
    [v / x] \Nat{n} &= \Nat{n} \\
    [v / x] x &= v \\
    [v / x] y &= y && \text{if } x \neq y \\
    [v / x] (e_1(e_2)) &= ([v / x] e_1)([v / x] e_2) \\
    [v / x] (e_1 + e_2) &= ([v / x] e_1) + ([v / x] e_2) \\
    [v / x] \Lam{y}{e} &= \Lam{y}{[v / x] e} && \text{if } x \neq y \\
    [v / x] \Lam{y}{e} &= \Lam{y}{e} && \text{if } x = y \\
    [v / x] \Filter{(p, a, g)}{e} &= \Filter{([v / x]p, a, g)}{[v / x] e} \\
    [v / x] \Residue{a}{g}{l}{e} &= \Residue{a}{g}{l}{[v / x] e} \\
    [v / x] \Fix{x}{e} &= \Fix{x}{e} \\
    [v / x] \Fix{y}{e} &= \Fix{y}{[v / x]e} && \text{if } x \neq y
  \end{aligned}
\]

\fbox{\([v / x] p = p'\)} \(p'\) can be obtained by substitution of \(v\) for
\(x\) in pattern \(p\).
\[
  \begin{aligned}
    [v / x] \$e &= \$e \\
    [v / x] \$v &= \$v \\
    [v / x] \Nat{n} &= \Nat{n} \\
    [v / x] y &= y && \text{if } x \neq y \\
    [v / x] (p_1(p_2)) &= ([v / x] p_1)([v / x] p_2) \\
    [v / x] (p_1 + p_2) &= ([v / x] p_1) + ([v / x] p_2) \\
    [v / x] \Lam{y}{e} &= \Lam{y}{[v / x] e} && \text{if } x \neq y \\
    [v / x] \Lam{y}{e} &= \Lam{y}{e} && \text{if } x = y \\
    [v / x] \Fix{x}{e} &= \Fix{x}{e} \\
    [v / x] \Fix{y}{e} &= \Fix{y}{[v / x]e} && \text{if } x \neq y
  \end{aligned}
\]

\jbox{\(\Decompose{e}{\mathcal{E}}{e'}\)} Expression \(e\) can be decomposed into context \(\mathcal{E}\) and redex \(e'\).
\begin{mathpar}
  \inferrule[D-Residue-T]{
    \Decompose{e}{\mathcal{E}}{e'}
  }{
    \Decompose{\Residue{a}{g}{l}{e}}{\Residue{a}{g}{l}{{\mathcal{E}}}}{e'}
  } \qquad
  \inferrule[D-Residue-E]{
    \Value{v}
  }{
    \Decompose{\Residue{a}{g}{l}{v}}{\circ}{\Residue{a}{g}{l}{v}}
  } \\
  \inferrule[D-Filter-T]{
    \Decompose{e}{\mathcal{E}}{e'}
  }{
    \Decompose{\Filter{f}{e}}{\Filter{f}{\mathcal{E}}}{e'}
  } \qquad
  \inferrule[D-Filter-E]{
    \Value{v}
  }{
    \Decompose{\Filter{f}{v}}{\circ}{\Filter{f}{v}}
  } \\
  \inferrule[D-Ap-L]{
    \Decompose{e_1}{\mathcal{E}_1}{e_1'}
  }{
    \Decompose{e_1(e_2)}{\mathcal{E}_1(e_2)}{e_1'}
  } \qquad
  \inferrule[D-Ap-R]{
    \Value{e_1} \\
    \Decompose{e_2}{\mathcal{E}_2}{e_2'}
  }{
    \Decompose{e_1(e_2)}{e_1(\mathcal{E}_2)}{e_2'}
  } \qquad
  \inferrule[D-Ap-E]{
    \Value{e_1} \\
    \Value{e_2}
  }{
    \Decompose{e_1(e_2)}{\circ}{e_1(e_2)}
  } \\
  \inferrule[D-Add-L]{
    \Decompose{e_1}{\mathcal{E}_1}{e_1'}
  }{
    \Decompose{e_1 + e_2}{\mathcal{E}_1 + e_2}{e_1'}
  } \qquad
  \inferrule[D-Add-R]{
    \Value{e_1} \\
    \Decompose{e_2}{\mathcal{E}_2}{e_2'}
  }{
    \Decompose{e_1 + e_2}{e_1 + \mathcal{E}_2}{e_2'}
  } \qquad
  \inferrule[D-Add-E]{
    \Value{e_1} \\
    \Value{e_2}
  }{
    \Decompose{e_1 + e_2}{\circ}{e_1 + e_2}
  } \\
  \inferrule[D-Fix-E]{
    \
  }{
    \Decompose{\Fix{x}{e}}{\circ}{\Fix{x}{e}}
  }
\end{mathpar}

\jbox{\(\Compose{e}{\mathcal{E}}{e'}\)} Expression \(e\) can be obtained by plugging \(e'\) into the mark of \(\mathcal{E}\).
\begin{mathpar}
  \inferrule[C-Top]{
  }{
    \Compose{e}{\circ}{e}
  } \\
  \inferrule[C-Ap-L]{
    \Compose{e_1}{\mathcal{E}_1}{e_1'}
  }{
    \Compose{e_1(e_2)}{\mathcal{E}_1(e_2)}{e_2'}
  } \qquad
  \inferrule[C-Ap-R]{
    \Compose{e_2}{\mathcal{E}}{e_2'}
  }{
    \Compose{e_1(e_2)}{e_1(\mathcal{E}_2)}{e_2'}
  } \\
  \inferrule[C-Add-L]{
    \Compose{e_1}{\mathcal{E}_1}{e_1'}
  }{
    \Compose{e_1 + e_2}{\mathcal{E}_1 + e_2}{e_1'}
  } \qquad
  \inferrule[C-Add-R]{
    \Compose{e_2}{\mathcal{E}_2}{e_2'}
  }{
    \Compose{e_1 + e_2}{e_1 + \mathcal{E}_2}{e_2'}
  } \\
  \inferrule[C-Filter]{
    \Compose{e}{\mathcal{E}}{e'}
  }{
    \Compose{\Filter{f}{e}}{\Filter{f}{\mathcal{E}}}{e'}
  } \qquad
  \inferrule[C-Residue]{
    \Compose{e}{\mathcal{E}}{e'}
  }{
    \Compose{\Residue{a}{g}{l}{e}}{\Residue{a}{g}{l}{\mathcal{E}}}{e'}
  }
\end{mathpar}


\end{document}